\newcommand{\be}{\begin{equation}}
\newcommand{\ee}{\end{equation}}
\newcommand{\bea}{\begin{eqnarray}}
\newcommand{\eea}{\end{eqnarray}}
\newcommand{\nn}{\nonumber}
\newcommand{\Appendix}[1]%
    {\renewcommand{\thesection}{Appendix~\Alph{section}:}%
         \section{#1}}%

\catcode`@=11
\long\def\@makecaption#1#2{
   \vskip 10pt
   \setbox\@tempboxa\hbox{{\small\bf #1.} \ {\small #2}}
   \ifdim \wd\@tempboxa >\hsize       
   {\small\bf #1.} \ {\small #2}\par  
   \else                              
        \hbox to\hsize{\hfil\box\@tempboxa\hfil}
   \fi}
\catcode`@=12

\catcode`@=11
\def\secteqno{\@addtoreset{equation}{section}%
\def\theequation{\thesection.\arabic{equation}}}
\def\endsecteqno{\def\theequation{\@ifundefined{chapter}%
{\arabic{equation}}{\thechapter.\arabic{equation}}}}
\newcounter{subequation}
\def\thesubequation{\alph{subequation}}
\def\sneqnarray{\stepcounter{equation}\let\@currentlabel=\theequation
\setcounter{subequation}{1}
\def\@eqnnum{{\rm (\theequation\thesubequation)}}
\global\@eqcnt\z@\tabskip\@centering\let\\=\@eqncr\let\@@eqncr=\@@sneqncr
$$\halign to \displaywidth\bgroup\@eqnsel\hskip\@centering
 $\displaystyle\tabskip\z@{##}$&\global\@eqcnt\@ne
 \hskip 2\arraycolsep \hfil${##}$\hfil
 &\global\@eqcnt\tw@ \hskip 2\arraycolsep
$\displaystyle\tabskip\z@{##}$\hfil
tabskip\@centering&\llap{##}\tabskip\z@\cr}
\def\endsneqnarray{\@@sneqncr\egroup $$\global\@ignoretrue}
\def\@@sneqncr{\let\@tempa\relax
   \ifcase\@eqcnt \def\@tempa{& & &}\or \def\@tempa{& &}
   \else \def\@tempa{&}\fi
     \@tempa \if@eqnsw\@eqnnum\stepcounter{subequation}\fi
     \global\@eqnswtrue\global\@eqcnt\z@\cr}
\def\nobiblabels{\def\@lbibitem[##1]##2{\@bibitem{##2}}}
\catcode`@=12

\def\beq{\begin{equation}}
\def\eeq{\end{equation}}
\def\bea{\begin{eqnarray}}
\def\eea{\end{eqnarray}}
\def\nn{\nonumber}

\newcommand{\eq}[1]{Eq.~\eqref{#1}}
\newcommand{\eqs}[2]{Eqs.~\eqref{#1} and \eqref{#2}}

\newcommand{\eqss}[2]{Eqs.~\eqref{#1}-\eqref{#2}}
\newcommand{\Sec}[1]{Sec.~\ref{#1}}

\def\la{\lambda} \def\lap{\lambda^{\prime}} \def\pa{\partial} \def\de{\delta}  \def\dag{\dagger}
   \def\Oc{{\rm O}} \def\S{{\rm S}}
\def\bnabla{{\bm \nabla}}
\def\bsigma{{\bm \sigma}}
\def\lQ{\Lambda_{\rm QCD}}

\documentclass[aps,12pt,prd,showpacs,amsmath,amssymb,preprintnumbers,superscriptaddress,nofootinbib,showkeys,preprint]{revtex4-1}

\usepackage[german,english]{babel}
\usepackage{hyperref}
\usepackage{amssymb}
\usepackage{amsmath}
\usepackage{bm}
\usepackage{braket}
\usepackage{slashed}
\usepackage{multirow}
\usepackage{epsfig}
\usepackage{epstopdf}
\usepackage{bbm}
\usepackage{soul}
\usepackage{xcolor}

\begin{document}

\title{Novel implementation of the multipole expansion to quarkonium hadronic transitions}

\author{Antonio Pineda and Jaume Tarr\'us Castell\`a}
\affiliation{Grup de F\'\i sica Te\`orica, Dept. F\'\i sica and IFAE-BIST, Universitat Aut\`onoma de Barcelona,\\ 
E-08193 Bellaterra (Barcelona), Spain}

\date{\today}

\begin{abstract}
We compute hadronic transitions between heavy quarkonium states with two, or one, pion/eta particles in the final state. We use the multipole expansion but not the twist expansion. The latter cannot be justified for the energy release of hadronic transitions between heavy quarkonium states with different principal quantum numbers. Instead, we use a counting based on the dimension of the interpolating field of the hybrid. This alternative counting allows us to still use chiral low-energy theorems to compute the pion production by local gluonic operators. We explore the phenomenological impact of this counting. Remarkably enough, for the two-pion transitions, we obtain the same predictions for the normalized differential decay rate as those obtained assuming the twist expansion. We implement this computational scheme using the hadronic representation of the effective theory potential NRQCD. We assume that the inverse Bohr radius of the heavy quarkonium is much larger than $\lQ$ but do not impose any constraint on the relative size of $\lQ$ and the typical kinetic energy of the bound state. 
\end{abstract}

\maketitle

\tableofcontents

\section{Introduction}

Hadronic transitions between heavy quarkonium states have been studied since the middle seventies. Particular attention has been devoted to the transitions with one or two pions in the final state. The relatively small energy of the outgoing pions makes the analysis of these transitions ideal for the implementation of (the nonlinear realization of) chiral symmetry. This was first implemented by Cahn and Brown in Ref.~\cite{Brown:1975dz}. Using current algebra, they parametrized the amplitude of the two-pion transition in the strict chiral limit. The observation of the small variation of the width distribution with respect to the helicity angle of one of the pions led them to an approximated simple one-parameter description of the decay width spectrum. The implementation of chiral symmetry using chiral Lagrangians and the incorporation of the leading light-quark mass corrections was made in Ref.~\cite{Mannel:1995jt}. This gave a good description of the width spectrum, and also  of the pion helicity angle distribution \cite{Yan:1998wv,Bai:1999mj}.

Alternatively, in Ref.~\cite{Gottfried:1977gp}, Gottfried realized that these transitions could be though as a two-step process: first a short-distance gluon emission by the heavy quarks, and then the hadronization of the gluons into the light-quark hadrons at a relatively long distance. This picture follows from the use of the multipole expansion of the gluon-heavy quark interaction. This allowed Gottfried to give selection rules and decay width rate estimates beyond the approaches based solely on chiral symmetry. The work of Ref.~\cite{Yan:1980uh} connected these nonlocal gluonic matrix elements with chiral symmetry by parametrizing them according to chiral symmetry. This was used to obtain constrains for higher angular momentum channels. 

At leading order (LO) in the multipole expansion, the intermediate heavy quarkonium state is in a color-octet configuration. In Ref.~\cite{Voloshin:1978hc}, Voloshin introduced an additional expansion consisting of an operator product expansion of the nonlocal heavy quarkonium color-octet two-point function. As a result, the transition amplitude can be written as a series of matrix elements of local operators, which we will refer to as a twist expansion. As before, these local matrix elements can be hadronized and parametrized using chiral symmetry but the implementation of the axial \cite{Gross:1979ur,Novikov:1979uy} and energy-momentum tensor anomalies \cite{Voloshin:1980zf,Novikov:1980fa,Chivukula:1989ds} constraints their general structure. Since then it has become quite customary to describe the two-pion transitions using the multipole, twist, and the chiral expansion. See, for instance, Ref.~\cite{Chen:1997zza}, where the local gluonic matrix elements were obtained up to next-to-leading order (NLO) in the chiral expansions. Unfortunately, this computational scheme is not a model independent derivation of QCD. The reason is that there is no kinematic regime where the twist expansion can be justified, as shown in Ref.~\cite{Luty:1993xf}, because the transfer energy between heavy quarkonium states in these transitions, $E$, is of order $m_Qv^2$, whereas the twist expansion requires that $E \ll m_Qv^2$, as well as $\lQ \ll mv^2$.
 
We want to retake this discussion within the context of effective field theories (EFTs), which allows us to make a systematic analysis of the scales involved in the problem. We will do the analysis using the weak-coupling version of potential NRQCD (pNRQCD)~\cite{Pineda:1997bj,Brambilla:1999xf}. This allows us to obtain the multipole expansion in a controlled way, since we assume that $m_Qv \gg \lQ$. The resulting EFT has the multipole expansion built in in the Lagrangian. The LO Lagrangian will also have spin symmetry, which follows from the heavy quark mass expansion. To simplify the problem further, we will also organize the computation within a $1/N_c$ expansion. If we hadronize this EFT, we obtain a Lagrangian in terms of the singlet (standard heavy quarkonium), hybrids and pion fields. $B/D$ mesons and possible tetraquarks are, \textit{a priori}, subleading in the large $N_c$, as we will discuss in more detail later in the paper. Therefore, we will not consider these degrees of freedom in this paper and their possible incorporation will be relegated to future work. As their effect could be important for hadronic transitions of states close or above to open flavor thresholds, we focus on states below threshold in this paper.   

We then write the most general hadronic representation of the weakly coupled pNRQCD Lagrangian made of the singlet, hybrids and pions fields in a combined expansion in the chiral counting and the multipole, $1/m_Q$ and $1/N_c$ expansions. It is formally possible to obtain the coefficients of the Lagrangian by matching to suitable Green functions in the pNRQCD theory in terms of quarks and gluons. Nevertheless, these coefficients endure a complicated relation with the elementary fields of the theory, and to determine them would require quite costly lattice simulations. Nevertheless, no all hope is lost. In Ref.~\cite{Brambilla:1999xf}, it was observed a correlation between the dimensionality of the interpolating operator of the hybrid/gluelump and its position in the spectrum. Though not equivalent, we hypothesize in this paper that there is also a correlation between the dimensionality of the interpolating operator and the strength of the interpolation with the hybrid, such that higher dimension operators are subleading (in this respect, it would also be interesting to study whether this approximation can be related with the analysis made in \cite{Fitzpatrick:2013twa}). We will see that such hypothesis plus the chiral low-energy theorems (generated by the axial and energy-momentum tensor anomalies) lead to the same predictions for the normalized differential decay rates of the two-pion transitions as using the twist expansion. Moreover, we do not need to impose extra conditions on $\lQ$, except the one we already imposed for the multipole expansion $m_Qv \gg \lQ$. In other words, our computational framework would still be valid even if $\lQ \gg m_Qv^2$. We explore the implications of this computational framework for a series of observables. Finally, we want to mention that the pure quarkonium hybrid sector of this theory has already been developed in Refs.~\cite{Berwein:2015vca,Brambilla:2017uyf,Oncala:2017hop,Brambilla:2018pyn} for small energy fluctuations (much smaller than the energy transitions we consider in this paper). 

We organize the paper as follows: in Sec.~\ref{hadpnrqcd}, we discuss the hadronization of the pNRQCD Lagrangian. In Sec.~\ref{ddws}, we apply it to describe the $Q\bar{Q}(2S)\to Q\bar{Q}(1S)\pi\pi$ transitions and compare the results for the decay width spectrum with experiment and the purely chiral description. Extending our Lagrangian beyond LO we use our approach to study one-pion transitions, $Q\bar{Q}(2S)\to Q\bar{Q}(1P)\pi$ in Sec.~\ref{2s1ptran}, and $Q\bar{Q}(2S)\to Q\bar{Q}(1S)\pi$ in Sec.~\ref{2s1s1pitran}. Certain uncertainties of our approach cancel out for specific ratios of the decay widths. We compute and study them in Sec~\ref{Sec:Ratios}. Finally, we give our conclusions in Sec.~\ref{conc}.

\section{Hadronization of the \texorpdfstring{\lowercase{p}}{p}NRQCD Lagrangian}\label{hadpnrqcd}

The pNRQCD Lagrangian at LO in $1/m_Q$ (except for the kinetic term) and at NLO in the multipole expansion reads
\begin{align}
L_{\rm pNRQCD} =& \int d^3R\Bigg\{\int d^3r \,\Bigl( {\rm Tr}\left[\S^{\dag}\left(i\partial_0-h_s\right)\S+ \Oc^{\dagger}\left(iD_0-h_o\right)\Oc\right] \nn \\
&
-\frac{1}{4} G_{\mu \nu}^a G^{\mu \nu\,a} + \sum^{n_f}_{i=1}\bar{q}_i (i\slashed{D} -m_i)q_i
 \nn
 \\
 &
 + g V_A ( r) {\rm Tr} \left[  {\rm O}^\dagger {\bf r} \cdot {\bf E} \,{\rm S}
+ {\rm S}^\dagger {\bf r} \cdot {\bf E} \,{\rm O} \right]
  + g \frac{V_B (r)}{2} {\rm Tr} \left[  {\rm O}^\dagger \left\{{\bf r} \cdot {\bf E} , {\rm O}\right\}\right]
\Biggr\}\,.
\label{pnrqcd1}
\end{align}
 $\S$ and $\Oc$ are the quark singlet and octet fields, respectively, normalized with respect to color as $S=S \bm{1}_c/\sqrt{N_c}$ and $\Oc=O^a T^a/\sqrt{T_F}$. They should be understood as functions of $t$, the relative coordinates $\bm{r}$, and the center of mass coordinates $\bm{R}$ of the heavy quarks. The trace should be understood as a double trace in color and spin. The singlet, octet and hybrid fields in the Lagrangians that appear in this paper are organized in $SU(2)$ spin multiplets. For instance, $S=\frac{1}{\sqrt{2}}({\bf S}\cdot {\bsigma}+S_{\eta}{\cal I})$. All the fields of the light degrees of freedom in \eq{pnrqcd1} are evaluated at $\bm{R}$ and~$t$; in particular, $G^{\mu \nu\,a}\equiv G^{\mu\nu\,a}(\bm{R},\,t)$, $q_i\equiv q_i(\bm{R},\,t)$, and  $iD_0 O\equiv i \partial_0O-g\left[A_0(\bm{R},\,t),O\right]$.   $h_s$ and $h_o$ are the singlet and octet Hamiltonian densities. They read as
\begin{align}
h_s=&-\frac{\bnabla^2_r}{m_Q}+V_s(r)\,,\\
h_o=&-\frac{\bnabla^2_r}{m_Q}+V_o(r)\,,
\end{align}
where $V_s(r)$ and $V_o(r)$ are computed in perturbation theory. Note that we have spin symmetry. Unless stated otherwise we will work in the isospin limit $m_i=\hat m \equiv \frac{m_u+m_d}{2}$ with $i=u,d$. At leading log~\cite{Pineda:2000gza}, and next to leading log \cite{Brambilla:2009bi} accuracy $V_A=1$.  

We next aim to construct the hadronic version of the above Lagrangian. We first need to characterize the hadronic degrees of freedom relevant to our case. We first work in the static limit.  In the short heavy-quark-antiquark distance limit the gluonic excitations can be characterized by the so-called gluelump operators. They organize themselves in irreducible representations of the $O(3) \otimes$C group. The LO Hamiltonian density in the $1/m_Q$ and multipole expansions corresponding to the Lagrangian in \eq{pnrqcd1} is given by
\begin{align}
H
&=\int d^3{\bf R}\int d^3{\bf r} \, {\rm Tr}\left[\S^{\dag}h_s\S+ \Oc^{\dagger}h_o\Oc\right]
\nn
\\
&+ \int d^3{\bf R} \left( \frac{1}{2} \left(\bm{E}^a\cdot\bm{E}^a+\bm{B}^a\cdot\bm{B}^a\right)
- \sum^{n_i}_{i=1}\bar{q}_i \,[ i  \bm{D}\cdot{\bm \gamma} -m_i]\, q_i \right)
\,.
\label{h-total}
\end{align} 
For later convenience, we will restrict the discussion to $L=1$ gluelump states. We define the gluelump operators, $G_{k}^{ia}$, as the color-octet gluonic operators that generate the eigenstates of $H$ in the presence of a local heavy-quark-antiquark octet source:
\begin{align}
H O^{a \dagger}({\bf R},{\bf r}) G^{ia}_{k}(\bm{R})|0\rangle = (V_o^{(0)}+\Lambda_{k}) O^{a \dagger}
({\bf R},{\bf r})G^{ia}_{k}(\bm{R})|0\rangle\,,
\end{align}
where $a$ is the color index, $k$ labels the quantum $J^{PC}$ numbers of the gluelump, and $i$ labels its vector components. At this stage, we do not have to make explicit the spin content of $O^a$. We normalize the gluelump operators as
\begin{align}
\langle 0|G_k^{ia\,\dagger}(\bm{R'})\,O^{a}(\bm{R'},{\bf r}')O^{b\,\dagger}(\bm{R},{\bf r})G_{k^{\prime}}^{jb}(\bm{R})|0\rangle=\de^{ij}\de_{kk^{\prime}}\delta({\bf R'}-{\bf R})\delta({\bf r'}-{\bf r})
\,.\label{glmpn}
\end{align}

Going beyond the LO in the multipole expansion the system is no longer spherically symmetric, instead  it is cylindrically symmetric around the heavy-quark-antiquark axis.\footnote{The symmetry group is $D_{\infty h}$, with P replaced by CP.} Representations of the cylindrical symmetry group can be constructed by projecting the gluelump operators on various directions with respect to the heavy-quark-antiquark axis. Therefore, we work with states with good transformation properties under the cylindrical symmetry group
\begin{align}
|{\bf R}, {\bf r};k,\,\lambda\rangle =P^i_{k\lambda} O^{a\,\dagger}\left(\bm{R},\bm{r}\right) G_{k}^{ia}(\bm{R})|0\rangle\,,
\label{eigen1}
\end{align}
where summation over index $i$ is implied. $P^i_{k\lambda}$ is a projector that acts onto the gluelump angular momentum and projects it into an eigenstate of $\mathbf{K}\cdot\hat{\mathbf{r}}$ (where $\mathbf{K}$ is the angular momentum operator for the gluelump) with eigenvalue $\lambda$.  

It is useful to project the pNRQCD Lagrangian onto the Fock subspace spanned by the $|{\bf R}, {\bf r};k,\,\lambda\rangle$ states
\begin{align}
\int d^3r d^3 R \, \sum_{k\lambda}|{\bf R}, {\bf r};k,\,\lambda\rangle \, \Psi_{k\lambda}(t,\,\bm{r},\,\bm{R})\,,
\label{boexpqcd}
\end{align}
where $\Psi_{k\lambda}(t,\,\bm{r},\,\bm{R})$ will represent the hybrid field in the hadronic version of the EFT. As we have already mentioned, the case of most interest to us is that of the spin $1$ gluelumps. The projectors for this case are
\begin{align}
P^{i}_{10}&=\hat{r}_0^i= \hat{r}^i\,,\label{pr10}\\
P^{i}_{1\pm 1}&=\hat{r}^i_{\pm}=\mp\left(\hat{\theta}^i\pm i\hat{\phi}^i\right)/\sqrt{2}\,,\label{pr11}
\end{align}
and the discrete symmetry transformations for the $\S$, $\Oc$, and $\Psi^i_{\kappa}$ fields are given in Table~\ref{Tab:discrete1}.
\begin{table}[ht!]
\begin{tabular}{c||c|c|c}
                           & {\rm P} & {\rm T} & {\rm C} \\ \hline\hline
$\S(t,\,\bm{r},\,\bm{R})$  & $-\S(t,\,-\bm{r},\,-\bm{R})$ & $\sigma_2\S(-t,\,\bm{r},\,\bm{R})\sigma_2$ & $\sigma_2\S^{\top}(t,\,-\bm{r},\,\bm{R})\sigma_2$ \\ \hline
$\Psi^i_{1^{+-}}(t,\,\bm{r},\,\bm{R})$ & $-\Psi^i_{1^{+-}}(t,\,-\bm{r},\,-\bm{R})$ & $-\sigma_2\Psi^i_{1^{+-}}(-t,\,\bm{r},\,\bm{R})\sigma_2$ & $-\sigma_2(\Psi^i_{1^{+-}})^{\top}(t,\,-\bm{r},\,\bm{R})\sigma_2$ \\ \hline
$\Psi^i_{1^{--}}(t,\,\bm{r},\,\bm{R})$ & $\Psi^i_{1^{--}}(t,\,-\bm{r},\,-\bm{R})$ & $\sigma_2\Psi^i_{1^{--}}(-t,\,\bm{r},\,\bm{R})\sigma_2$ & $-\sigma_2(\Psi^i_{1^{--}})^{\top}(t,\,-\bm{r},\,\bm{R})\sigma_2$ \\ \hline
$\bm{E}(t,\,\bm{R})$      & $-\bm{E}(t,\,-\bm{R})$     & $\bm{E}(-t,\,\bm{R})$          & $-\bm{E}^{\top}(t,\,\bm{R})$ \\ \hline
$\bm{B}(t,\,\bm{R})$      & $\bm{B}(t,\,-\bm{R})$      & $-\bm{B}(-t,\,\bm{R})$         & $-\bm{B}^{\top}(t,\,\bm{R})$ \\ \hline
$\hat{\bm{r}}_{\la}$    & $-\hat{\bm{r}}_{-\la}$  & $\hat{\bm{r}}^{*}_{\la}$    & $-\hat{\bm{r}}_{-\la}$ \\ 
\end{tabular}
\caption{Transformation properties of the heavy quarkonium and gluonic fields, and the projection vectors, under discrete symmetries. The octet field $\Oc$ has the same transformation properties as $\S$. The $\Psi^i_{k}$ transform as $\Oc$ combined with the $k^{PC}$ of the gluelump. The transformations of the projected fields can be obtained by further adding those of the projection vectors; however, these are actually not relevant for the construction of the Lagrangian since the projection vectors always appear in pairs, one explicit in the operator and another implicit in $\Psi_{k\lambda}=\hat{\bm{r}}_{\la}\cdot{\bf \Psi}_{k}$. For this reason, we give the transformation properties of the unprojected fields $\Psi^i_{k}$. Note that the difference in the transformation of $\hat{\bm{r}}_{\la}$ with respect to Ref.~\cite{Berwein:2015vca} are due to the different definition of $\hat{\bm{r}}_+$.}
\label{Tab:discrete1}
\end{table}

Besides the singlet and hybrid fields, we will incorporate pions to our Lagrangian. As a basic building block for the Goldstone bosons, we use the unitary matrix $U(t,\,\bm{R})$, which [for $SU(3)$] may be taken as
\be
U = e^{i\Phi /F}\,, \quad 
\Phi=\left( \begin{array}{ccc}
\pi^0+\frac{\eta}{\sqrt{3}} & \sqrt{2} \pi^+ & \sqrt{2} K^+\\
\sqrt{2} \pi^-              & -\pi^0+\frac{\eta}{\sqrt{3}} & \sqrt{2} K^0\\
\sqrt{2} K^-                & \sqrt{2} \bar{K}^0           & -\frac{2}{\sqrt{3}}\eta \\
\end{array}
\right)\,,
\label{pions}
\ee
although final results for observable quantities do not depend on this specific choice. $F=92.419$ MeV is the pion decay constant. Under chiral symmetry, $U$ transforms as $U\stackrel{g}{\to}RUL$, where $R\in SU_R(N)$ and $L\in SU_L(N)$. Related useful matrices are $u$, defined from $u^2=U$, and
\begin{align}
u_{\mu}=&i\left(u^{\dag}(\partial_{\mu}-i r_{\mu}))u-u(\partial_{\mu}-i l_{\mu})u^{\dag}\right)\,, \\
\chi_{\pm}=&u^{\dag}\chi u^{\dag}\pm u \chi^{\dag} u\,,
\end{align}
where $\chi =2B {\rm diag}(\hat{m},\hat{m},m_s)$, with $B$ being related to the vacuum quark condensate. In the isospin limit, the pion mass is $m_{\pi}^2 = 2 B\hat{m}$. The transformation properties can be found in Table~\ref{Tab:discrete2}.

\begin{table}
\begin{tabular}{c||c|c|c|c}
    & $u$        & $u_{\mu}$        & $\chi_{\pm}$        & $D_{\mu}$        \\ \hline\hline
P   & $u^{\dag}$ & $-u^{\mu}$       & $\pm\chi_{\pm}$     & $D^{\mu}$        \\ \hline
C   & $u^{\top}$ & $u^{\top}_{\mu}$ & $\chi^{\top}_{\pm}$ & $D^{\top}_{\mu}$ \\ \hline
T   & $u$        & $u^{\mu}$        & $\chi_{\pm}$        & $D^{\mu}$        \\ \hline
h.c & $u^{\dag}$ & $u_{\mu}$        & $\pm\chi_{\pm}$     & $D_{\mu}$        \\ 
\end{tabular}
\caption{Transformation properties of the basic chiral building blocks under discrete symmetries.}
\label{Tab:discrete2}
\end{table}

We now construct the hadronic Lagrangian. It is fixed by the degrees of freedom, the symmetries, and the parameter expansions we have: $1/N_c$, $r$, $E$, $1/m_Q$, $m_i$. We emphasize that, at this level, we do not integrate out extra degrees of freedom when going to the hadronic representation  of \eq{pnrqcd1}. Therefore, it is not a different EFT but the very same pNRQCD, including the same degrees of freedom and scales. Instead what we do is to write the most general Lagrangian consistent with the symmetries made out of the heavy quarkonium, hybrids, and pions. We write this Lagrangian at LO in the chiral counting and the $1/m_Q$ expansion (except for the kinetic term) and at NLO in the multipole expansion. For the interaction between heavy quarkonium, hybrids, and pions, we also incorporate the large $N_c$ expansion, and consider only the leading terms in it. Strictly speaking, one should also include glueballs, which may interact with the hybrids at LO in the multipole expansion, but at NLO in the $1/N_c$ expansion. Such effects would be subleading in our computation of the decays. Therefore, we neglect them. The hadronic version of the pNRQCD Lagrangian projected onto the subspace of \eq{boexpqcd} reads 
\begin{align}
&
L_{pNRQCD}^{\rm had} = \int d^3Rd^3r \, {\rm Tr}\Bigl[S^{\dag}\left(i\partial_t-V_s(r)+\frac{\bnabla^2_r}{m_Q}\right)\S+\frac{F^2}{4}\left(\langle u_{\mu}u^{\mu}\rangle+\langle\chi_+\rangle\right) \nn \\
&+\sum_{k^{PC}=1^{+-},\,1^{--}} \sum_{\la\lap}\Psi^{\dagger}_{k\lambda} \biggl\{(i\partial_t - (V_o^{(0)}+\Lambda_{k}))\delta_{\lambda\lambda'}+\hat{\bm{r}}^{i\dag}_{\la}\frac{\bnabla^2_r}{m_Q}\hat{\bm{r}}^i_{\lap}\biggr\}\Psi_{k\lap} \nn\\
&+\left(\bm{r}\cdot \hat{\bm{r}}_{\la}S^{\dag}\Psi_{1^{--}\la}+\text{h.c}\right)t^{(r1^{--})}+{\bf r} \cdot \delta {\bf {\cal L}}(\Psi_{k\la},\Psi_{k'\la'})
\nn\\
&+\left(\bm{r}\cdot \hat{\bm{r}}_{\la}S^{\dag}\Psi_{1^{--}\la}+\text{h.c}\right)\left(t^{(r1^{--})}_{d0}F^2\langle u_0u_0\rangle+t^{(r1^{--})}_{di}F^2\langle u_i u^i\rangle+t^{(r1^{--})}_m F^2\langle\chi_+\rangle\right)
\Bigr]
\,.
\label{bolag2}
\end{align}
Note that for each hybrid channel one should also include the excitations. The fields $S$ and $\Psi_{k\lambda}$ should be understood as depending on $t$, $\bm{r}$, and $\bm{R}$. The pion fields depend on $t$ and $\bm{R}$. Note that Tr[] now only stands for the trace over spin indices. $\langle A \rangle$ stands for the trace of $A$ in the isospin index.

Since at NLO in the multipole expansion the singlet can only mix with $k=1$ gluelumps, we only include hybrid states that can be generated by such gluelumps in the Lagrangian. According to the operator analysis of Table~II in Ref.~\cite{Brambilla:1999xf}, they correspond to $\Sigma_g^{+'}$, $\Pi_g$, $\Sigma_u^-$, $\Pi_u$, and associated excitations with bigger gluelump masses. The mixing between hybrids at NLO in the multipole expansion is more complicated. It is encoded in ${\bf r} \cdot \delta {\bf {\cal L}}(\Psi_{k\la},\Psi_{k'\la'})$, where $\delta {\bf {\cal L}}(\Psi_{k\la},\Psi_{k'\la'})$ is bilinear in the hybrid fields and transforms as a $1^{--}$ vector. Note that in general $k\not=1$ states may contribute to this term. Fortunately, we will not need the details of this interaction for the analysis of this paper. Note also that at NLO in the multipole expansion we do not have hybrid bound states but plane waves. We will need to iterate the ${\cal O}(r)$ vertices to obtain bound states in the hybrid sector.

The last line in \eq{bolag2} encodes the interaction of the hybrids with the singlet and pions at NLO in the multipole expansion and at the leading nonvanishing order in the large $N_c$ and chiral expansions. The interaction with two pions scales with $N_c$ as $1/N_c^2$. One may consider the incorporation of subleading operators in the chiral counting but still leading in the $1/N_c$ expansion. Since for the process we consider in this paper, the typical energy of the pions will be of ${\cal O}(mv^2)$ or smaller (which in general we will consider them to be smaller than $\lQ$); this will not be necessary.  This is also the reason we do not include interactions of the pions with the hybrids that are not suppressed by the multipole expansion. The contribution of these operators to heavy quarkonium hadronic transitions would be suppressed by a power of $mv^2/\lQ$ compared with the contributions considered in this paper.

In the above discussion we have not included tetraquarks (like the $Z_b$ or $Z_c$) , nor $Q\bar q$-$\bar Q q$ states, in the physical spectrum. One may wonder whether we should do so. Here we will guide our discussion by the $1/N_c$ counting. The inclusion of tetraquarks is delicate. They are not stable in the large $N_c$. If the decay width grows like $N_c$, then $\Gamma \gg E$ and diagrams with intermediate tetraquarks become effectively local and $1/N_c$ suppressed relative to the accuracy for the physical processes we consider in this paper. The same happens for the $Q\bar q$-$\bar Q q$ loops. The suppression of tetraquark effects can only be bypassed if, for some reason, $\Gamma$ is much smaller than what is expected by the $N_c$ counting. This could happen if the channels that would contribute at leading order in $N_c$ are closed because they are below threshold. This is the situation discussed in Ref.~\cite{Weinberg:2013cfa}. Then, indeed, the scaling in $N_c$ of $\Gamma$ is, at most, of order $1/N_c$. However, we are not in this situation. For instance, $Z_b$ states are above threshold, albeit very close to it. Therefore, there is little phase space free, effectively producing that $\Gamma$ is very small. In this scenario the mixing with $Q\bar q$-$\bar Q q$ loops is expected to be very large and it does not make much sense to consider one without the other. In this respect, it is interesting to note that in Refs.~\cite{Chen:2015jgl,Chen:2016mjn} it has been advocated that tetraquarks and $Q\bar q$-$\bar Q q$ loops may play an important role for some observables like the $Q\bar Q(3S/4S) \rightarrow  Q\bar Q(1S)$ decays. Nevertheless, in this paper, we want to specifically study the effect associated to the inclusion of hybrids within a context where the multipole expansion can be applied. Therefore, the possible incorporation of tetraquark and $Q\bar q$-$\bar Q q$ states will be postponed to future work. 

In this paper, we will generally consider that the quarkonium binding energies fulfill $m_Qv^2\ll \lQ$. This allows us to write the static singlet potential up to ${\cal O}(r^2)$ as
\begin{align}
V_{\Sigma^+_g}(r)&=\lim_{t \rightarrow \infty}\frac{i}{t}\ln\left(\langle 0|S(t,\,\bm{r},\bm{R'})S^{\dag}(0,\,\bm{r},\bm{R})|0\rangle\right)=V^{(0)}_s(r)+b_{\Sigma^+_g} r^2+\cdots .
\label{VSinglet}
\end{align}
The coefficient $b_{\Sigma^+_g}$ can actually be determined in terms of the coefficients $t^{(r1^{--})}$ from the Lagrangian in \eq{bolag2}, since ${\cal O}(r^2)$ terms in the Lagrangian do not contribute to the static energies of the singlet.

In the case of hybrid bound states, $m_Qv^2\ll \lQ$ is also the natural hierarchy between the scales since the bound states are small energy fluctuation around the minima of the hybrid static energies. Therefore, we can also generically write 
\begin{align}
V_{k\la\lap}(r)&=\lim_{t \rightarrow \infty}\frac{i}{t}\ln\left(\langle 0| O^{a}\left(t,\bm{r},\bm{R'}\right)P^{\dag}_{k\la}\cdot G_{k}^{a\dag}(t,\bm{R}') O^{a\,\dag}\left(0,\bm{r},\bm{R}\right)P_{k\lap}\cdot G_{k}^{a}(0,\bm{R})|0\rangle\right)\nn\\
&=\left(\Lambda_{k}+V^{(0)}_o+b_{k\la} r^2+\cdots\right)\de_{\la\lap}+\mathcal{O}(1/m_Q)\, .\label{Vkappa}
\end{align}
Nevertheless, in this case, we cannot guaranty that ${\cal O}(r^2)$ terms in the Lagrangian will not contribute to the hybrid potential. Therefore, we cannot determine $b_{k\la}$ from the ${\cal O}(r)$ coefficients of the hadronic Lagrangian alone. We will determine this coefficient by fitting the potential to the lattice data for the static energies.

The LO singlet and hybrid spectrum and wave functions will be obtained from solving the Schr\"odinger equation with different variants of the singlet and hybrid potential based on Eqs.~\eqref{VSinglet} and \eqref{Vkappa}. The details of these solutions can be found in Appendix~\ref{AppStates}.

Let us now discuss the vertices of the hadronic Lagrangian. To obtain the coupling in the hadronic Lagrangian we match equivalent correlation functions computed in both representations of the theory. The equalities are obtained in the static limit. The singlet-hybrid mixing term in the hadronic theory is given by
\begin{align}
\langle 0|\Psi_{\la}(t,\,\bm{r}',\,\bm{R}')S^{\dag}(0,\,\bm{r},\,\bm{R})|0 \rangle_{\rm amp.}=i \bm{r}\cdot \hat{\bm{r}}^{\dag}_{\la}t^{(r1^{--})}
\,,
\label{s1mmmixa}
\end{align}
which is matched to the following correlator of pNRQCD in term of quarks and gluons:
\begin{align}
&Z_E\langle 0|\hat{\bm{r}}^{\dag}_{\la}\cdot \bm{E}^{a\dag}(t,\bm{R}')O^{a}\left(t,\,\bm{r}',\,\bm{R}'\right)S^{\dag}(0,\,\bm{r},\,\bm{R})|0\rangle_{\rm amp.}=i\sqrt{\frac{T_F}{N_c}}\hat{\bm{r}}^{\dag i}_{\la}\bm{r}^j \langle 0|G_{1^{--}}^{a i\dag}(\bm{R})g\bm{E}^{aj}(\bm{R})|0\rangle
\,,\label{s1mmmixb}
\end{align}
where \textit{amp.} signals that only amputated contributions are considered [overall $\delta({\bf r}'-{\bf r})$ are also factored out]. To evaluate the matrix element, we consider the interpolating field for $G_{1^{--}}^{a}$ to be given by a sum of all possible local gluonic operators with the same quantum numbers,
\begin{align}
G_{1^{--}}^{a}=Z^{-1/2}_E g\bm{E}^a+Z^{-1/2}_{D\times B}\left(\bm{D}\times g \bm{B}\right)^a+\cdots\label{1mmgle}
\,.
\end{align}
Now, we then hypothesize that there is a correlation between the dimensionality of the interpolating operator and the strength of the interpolation with the hybrid, such that higher dimension operators are subleading, so the series can be truncated at LO. Though not rigorous, one may consider counting $Z_O^{-1/2}\sim \Lambda^{1-{\rm dim}(O)}_{\rm QCD}$, where dim(O) is the dimension of the operator O, and the size of the operator to be given by $\sim \left(mv^2\right)^{{\rm dim}(O)}$, so the series can be truncated at LO up to corrections of $\mathcal{O}(mv^2/\Lambda_{\rm QCD})$. Using this truncation and \eq{glmpn} in \eq{s1mmmixb}, we obtain
\begin{align}
t^{(r1^{--})}=\sqrt{\frac{T_F Z_E}{N_c}}\,.
\end{align}

The operators with an even number of pions in \eq{bolag2} are matched in a similar way. In the hadronic pNRQCD,
\begin{align}
&\int d^4x_+d^4x_-e^{i p_+\cdot x_+}e^{ip_-\cdot x_-}\langle 0|\pi^+(x_+)\pi^-(x_-)S(t,\,\bm{r},\,\bm{R})\Psi^{\dag}_{1^{--}\la}(0,\,\bm{r},\,\bm{R})|0\rangle_{\rm amp.}\nn\\
&=i4\bm{r}\cdot \hat{\bm{r}}_{\la}\left(-t_{d0}^{(r1^{--})}p^0_+p^0_-+t_{di}^{(r1^{--})}\bm{p}_+\cdot\bm{p}_--t_{m}^{(r1^{--})}m^2_{\pi}\right)\label{2pma}\,,
\end{align}
and the corresponding correlator in the partonic pNRQCD reads
\begin{align}
&\langle \pi^+(p_+)\pi^-(p_-)|S(t,\,\bm{r},\,\bm{R})\hat{\bm{r}}_{\la}\cdot G^a_{1^{--}}(0,\,\bm{R}) O^{a\dag}(0,\,\bm{r},\,\bm{R})|0\rangle_{\rm amp.}\nn \\
&=ig\sqrt{\frac{T_F}{N_c}}\hat{\bm{r}}_{\la}\cdot\bm{r}\langle \pi^+(p_+)\pi^-(p_-)|\bm{E}(\bm{R})\cdot G_{1^{--}}(\bm{R})|0\rangle\nn\\
&=\frac{i}{3}\sqrt{\frac{T_F}{N_cZ_E}}\hat{\bm{r}}_{\la}\cdot\bm{r}\frac{8\pi^2}{\beta_0}\left(\left(2-\frac{9}{2}\kappa\right)p^0_+p^0_--\left(2+\frac{3}{2}\kappa\right)\bm{p}_+\cdot\bm{p}_-+3m^2_{\pi}\right)\,,\label{2pmb}
\end{align}
where in the last step we use \eq{1mmgle} truncated at the first term and hadronize the resulting gluonic matrix element using \eq{bbh2p}, which uses the anomaly relation of the energy-momentum tensor of QCD~\cite{Voloshin:2007dx,Novikov:1980fa}. The derivation of this equation is reviewed in Appendix~\ref{Sec:anomaly}.

Comparing Eqs.~\eqref{2pma} and \eqref{2pmb}, we obtain
\begin{align}
&t_{d0}^{(r1^{--})}=-\frac{2\pi^2}{3\beta_0}\sqrt{\frac{T_F}{N_cZ_E}}\left(2-\frac{9}{2}\kappa\right)\,,\label{tre0}\\ 
&t_{di}^{(r1^{--})}=-\frac{2\pi^2}{3\beta_0}\sqrt{\frac{T_F}{N_cZ_E}}\left(2+\frac{3}{2}\kappa\right)\,,\label{trei}\\
&t_{m}^{(r1^{--})}=-\frac{2\pi^2}{3\beta_0}\sqrt{\frac{T_F}{N_cZ_E}}3\,.\label{trem}
\end{align}

\section{\texorpdfstring{$\bar Q Q(2S) \rightarrow \bar Q Q(1S) \pi \pi$}{QbarQ(2S)->QbarQ(1S)pipi} hadronic transitions} \label{ddws}

The two-pion quarkonium transitions are well described using chiral perturbation theory and spin symmetry. The most general amplitude for the two-pion transitions at ${\cal O}(p^2)$ in the chiral counting reads 
\begin{align}
\mathcal{A}_{\chi}&=-a_1 p^0_+p^0_-+a_2 \bm{p}_+\cdot \bm{p}_--a_3m^2_{\pi}
\,,
\label{2pgp1}
\end{align}
where $p_+$, $p_-$ are the momentum of the $\pi^+$ and $\pi^-$, respectively. The coefficients $a_i$ can be thought as linear combinations of Wilson coefficients of an effective chiral Lagrangian made only by heavy quarkonium and pions. Such effective Lagrangian can be found in Eqs.~(4) and (6) of Ref.~\cite{Mannel:1995jt}. Imposing heavy quark spin symmetry sets the $g_2$ low-energy constant of Ref.~\cite{Mannel:1995jt} to zero. 

 \begin{figure}[ht!]
   \centerline{\includegraphics[width=.8\textwidth]{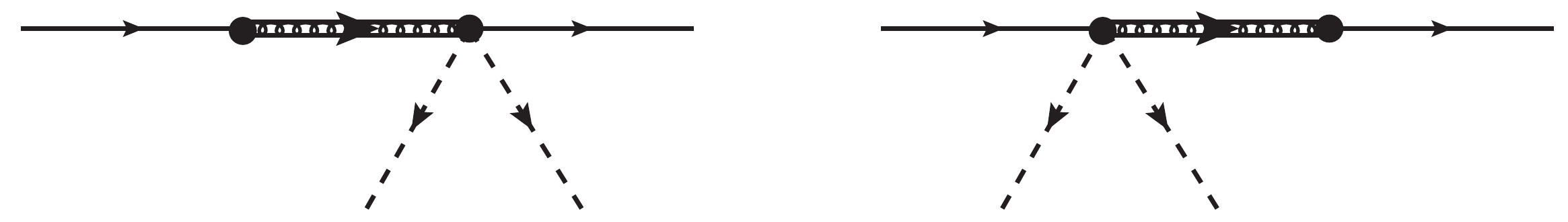}}
	\caption{Single and dashed lines represent quarkonia and pions. The double line with a curly line inside represents hybrid states.}
	\label{tptf}
 \end{figure}

Alternatively, we can use the computational scheme developed in this paper. Using \eq{bolag2}, we can compute the leading nonvanishing contribution to the two-pion transitions between quarkonium states. The diagrams are drawn in Fig.~\ref{tptf}. One vertex mixes singlet and hybrid fields whereas the other produces a two-pion emission vertex. The amplitude for this process reads
\begin{align}
i\mathcal{A}&= i\beta^{(n^{\prime}n)}_{r,Q} t^{(r1^{--})}\left(t_{d0}^{(r1^{--})}p^0_+p^0_--t_{di}^{(r1^{--})}\bm{p}_+\cdot\bm{p}_-+t_{m}^{(r1^{--})}m^2_{\pi}\right)\,,\label{qumqum2p}
\end{align}
with
\begin{align}
&\beta^{(n^{\prime}n)}_{r,Q}=\sum_m \langle S_{n^{\prime}} | 
\hat{\bm{r}}^{\dag}_{\la}\cdot\bm{r}|\Psi_m\rangle
\left(\frac{1}{m_{n}-m_m}+\frac{1}{m_{n'}-m_m}\right)\langle\Psi_m|
\hat{\bm{r}}^{\dag}_{\la}\cdot\bm{r}|S_n\rangle
\,.\label{betarc21}
\end{align}
 The index $m$ sums over all states solution of the Schr\"odinger equation and also for the gluelump excitations. For the second term in the brakets of \eq{qumqum2p} (corresponding to the right diagram in Fig.~\ref{tptf}), we have used that $m_{n}=m_{n^{\prime}}+p^0_{+}+p^0_{-}$. It is remarkable that the normalization factors $Z_E$ cancels out, which allows us to completely evaluate \eq{qumqum2p} except for the parameter $\kappa$ that appears in the couplings in Eqs.~\eqref{tre0}-\eqref{trem} and has its origin in the hadronization of the gluonic operator. In the next section we will fix $\kappa$ by fitting the normalized differential decay width spectrum.

For the case at hand, the quarkonia states are $n'=1^3S_1$ and $n=2^3S_1$. The intermediate hybrid has $1^{--}$ quantum numbers. Expressions for the hybrid-quarkonia matrix elements that appear in \eq{betarc21} are given in Appendix~\ref{app:matelem}. The wave functions of the bound states that appear in those matrix elements can be found in Appendix \ref{AppStates}. The matrix elements in \eq{betarc21} are diagonal in the spin state of the heavy quarks. Thus, only spin triplet hybrid states are allowed as intermediate states. Furthermore, the matrix elements in \eq{betarc21} only receive contribution from the $\lambda=0$ component of the hybrid state, which leads to the selection rule $l=\ell$ for the angular momentum eigenvalues. Therefore, the only possible hybrid intermediate states correspond to $m^3\mathcal{S}_1$ states. 

Equation~\eqref{qumqum2p} yields the following prediction for the chiral parameters $a_i$:
\begin{align}
&a_1=-\frac{8\pi^2T_F}{3\beta_0 N_c}\beta^{(12)}_{r,Q}\left(2-\frac{9}{2}\kappa\right)\,,\label{c21a1}\\
&a_2=-\frac{8\pi^2T_F}{3\beta_0 N_c}\beta^{(12)}_{r,Q}\left(2+\frac{3}{2}\kappa\right)\,,\label{c21a2}\\
&a_3=-\frac{8\pi^2T_F}{\beta_0 N_c}\beta^{(12)}_{r,Q}\,.\label{c21a3}
\end{align}

We now confront the above predictions with experiment. Since the experimental data on $\frac{d\Gamma}{d m_{\pi\pi}}$ (where $m_{\pi\pi}=(p_++p_-)^2$ is the dipion invariant mass) are normalized to an unknown constant, it is convenient to fit the theoretical expressions to the normalized differential decay width 
\be
\frac{1}{\Gamma}\frac{d\Gamma}{d m_{\pi\pi}}
\,.
\ee 
As we will see there is also a strong theoretical motivation to consider this ratio. This object is only sensitive to ratios of the theory parameters. The overall normalization can be obtained from the total decay width. Therefore, in the following subsections, we fit the normalized differential decay width and total decay width using the formulas discussed above. Detailed formulas can be found in Appendix~\ref{ddwap}. Here we will analyze the transitions $\psi(2S)\to J/\Psi \pi^+\pi^-$ and $\Upsilon(2S)\to \Upsilon(1S) \pi^+\pi^-$. The experimental data for the former are taken from Ref.~\cite{Aaboud:2016vzw}, and for the latter from Ref.~\cite{Alexander:1998dq}.\footnote{In the first version of this paper the more recent data from Ref.~\cite{Guido:2017cts} was used. However this data appears to have some problems. This reflects in an strange behavior of the angular distributions. We acknowledge conversations on this issue with the authors.} The total decay widths are taken from Ref.~\cite{Tanabashi:2018oca}.

\subsection{Line-shape analysis}
\label{Sec:Line-shape}
\begin{figure}[ht!]
\includegraphics[width=.6\textwidth]{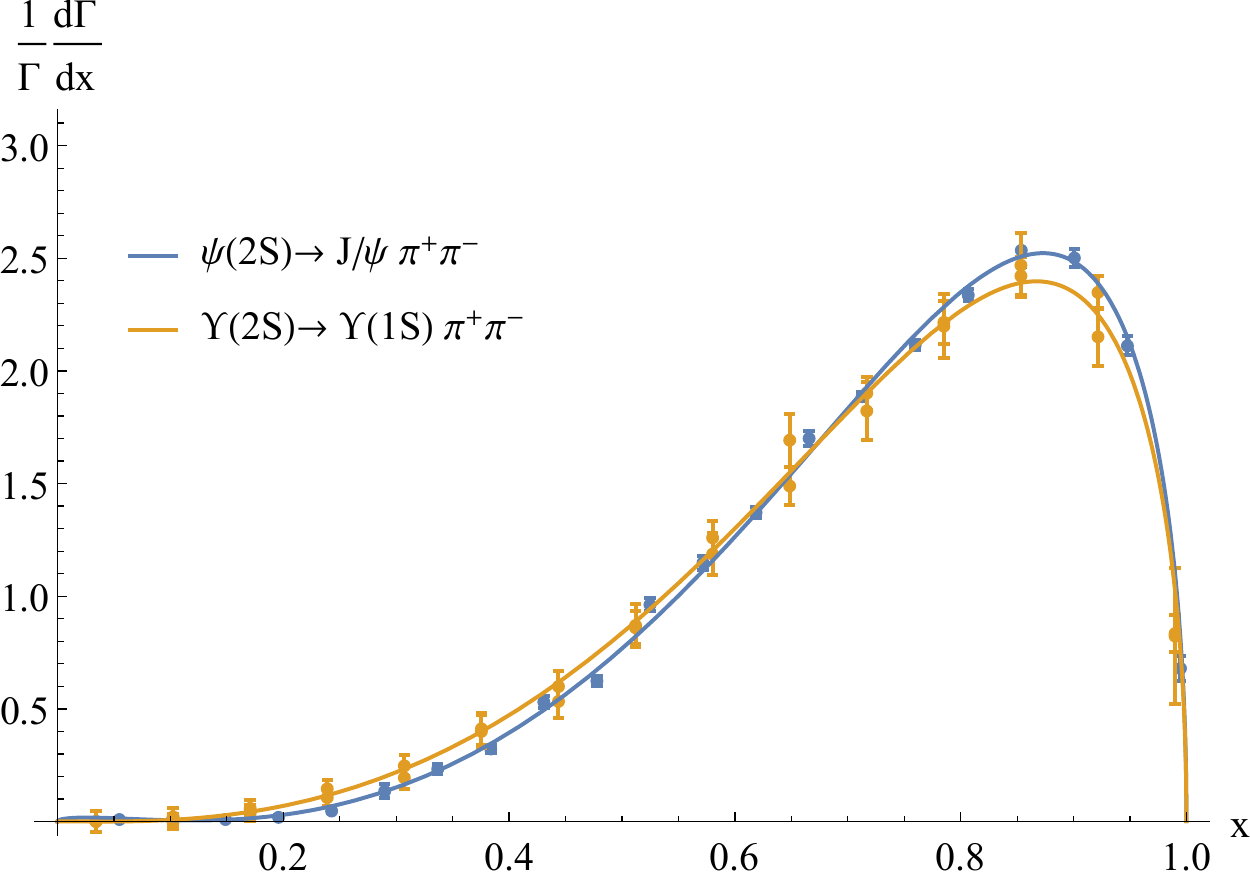}\\
\caption{Plot of the normalized differential decay width spectrum. The dots are the experimental data for $\psi(2S)\to J/\Psi \pi^+\pi^-$~\cite{Aaboud:2016vzw} and $\Upsilon(2S)\to \Upsilon(1S) \pi^+\pi^-$~\cite{Alexander:1998dq} in blue and yellow respectively. In the same color scheme, the continuous lines are the fits of the theoretical expression obtained from the amplitude in \eq{2pgp1} computed from an EFT incorporating chiral and spin symmetry. The variable $x$ is defined as $x=\frac{m_{\pi\pi}-2m_{\pi}}{m_{2S}-m_{1S}-2m_{\pi}}$.}
\label{cheft}
\end{figure}

\begin{figure}[ht!]
\begin{tabular}{cc}
\includegraphics[width=.49\textwidth]{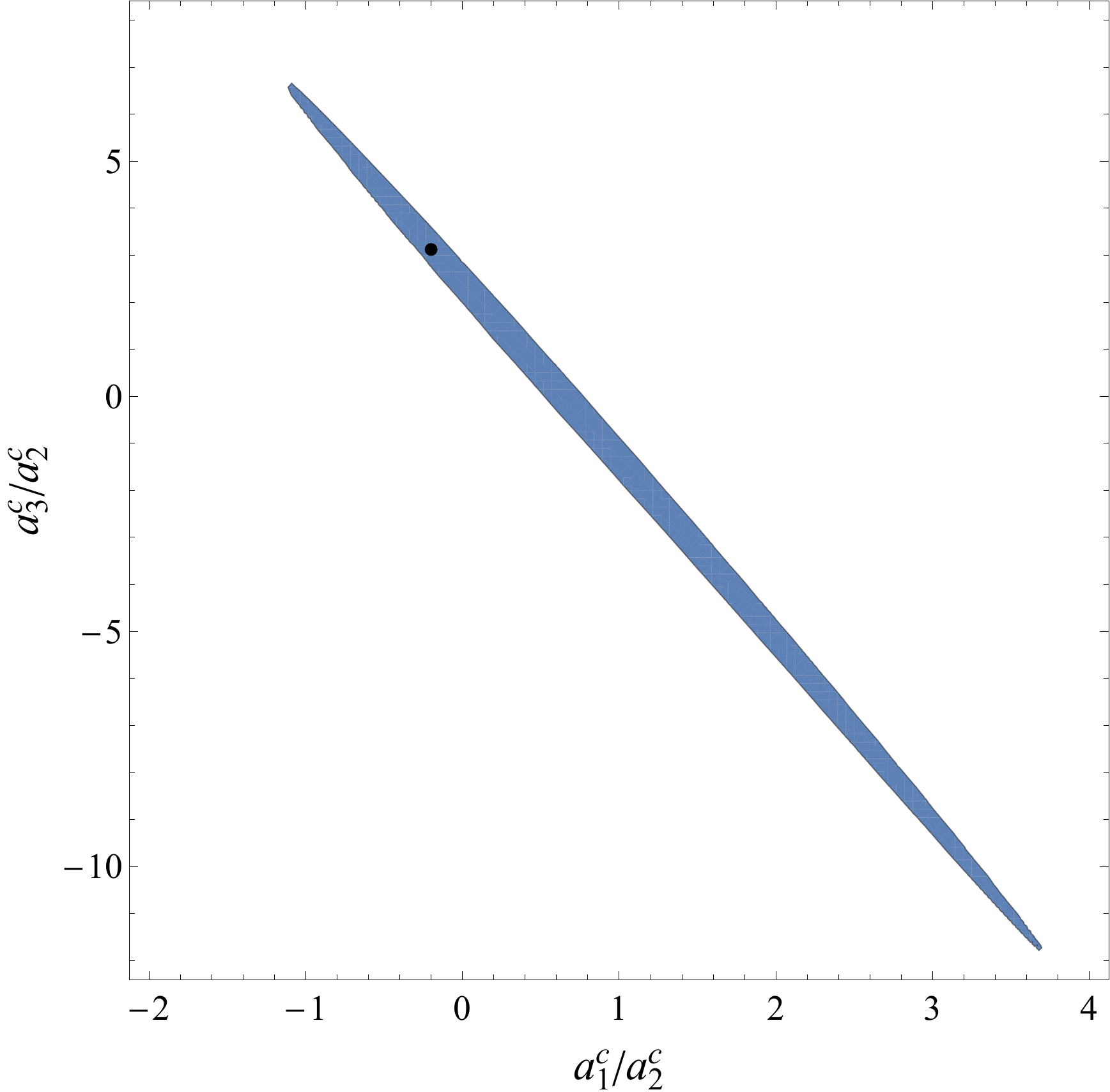} & \includegraphics[width=.49\textwidth]{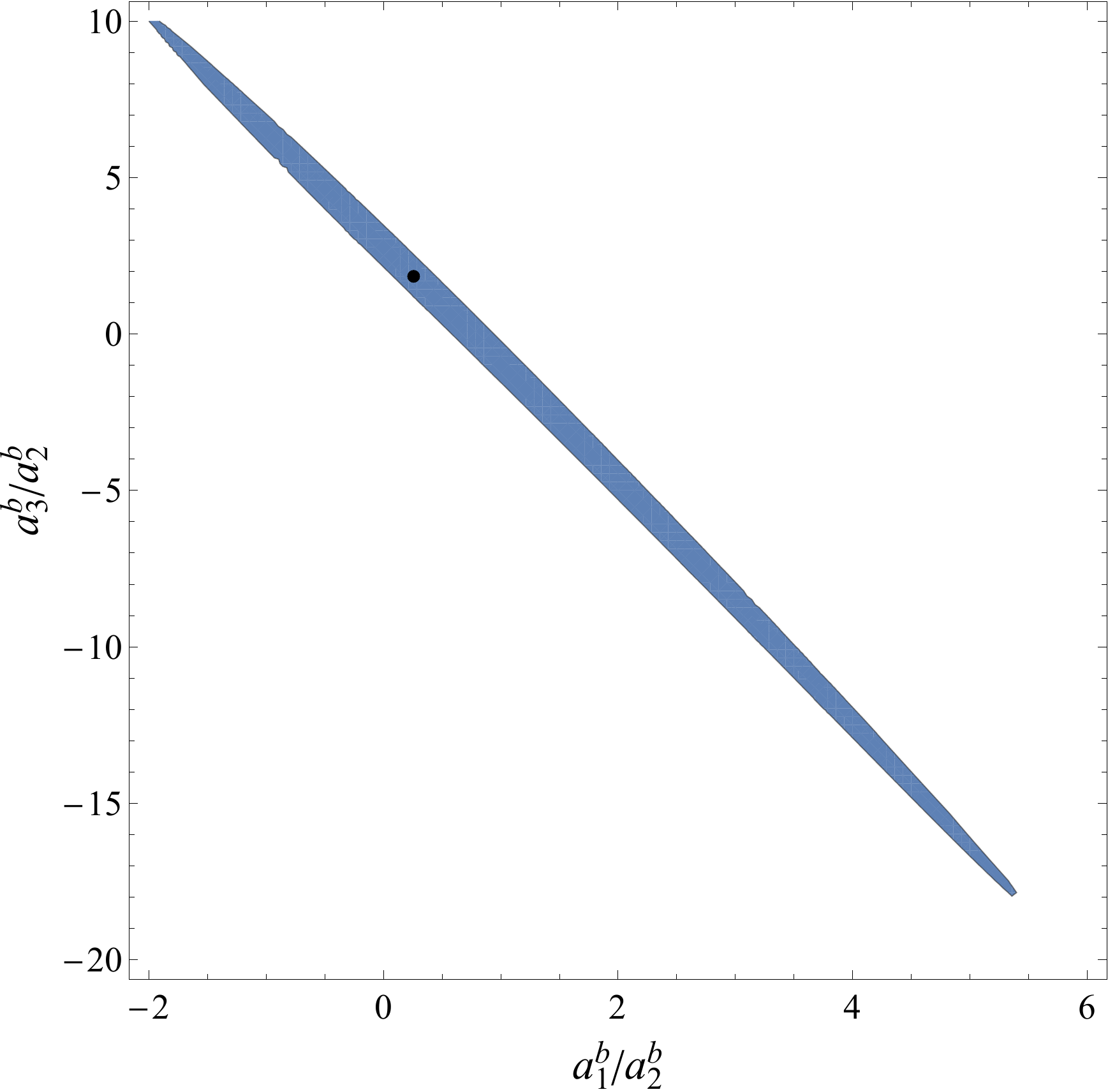} \\
\end{tabular}
\caption{Correlation between the parameters of the fit of the normalized decay width spectrum. Results for charmonium and bottomonium transitions in the left and right sides, respectively. The blue areas indicate the parametric space regions with $\chi^2_{\rm d.o.f}\leq 1$, and the black dots the best fit whose values are given in Eqs.~\eqref{fchc} and \eqref{fchb}.}
\label{corrplot}
\end{figure}

 The chiral fit to $\frac{1}{\Gamma}\frac{d\Gamma}{d m_{\pi\pi}}$ produces the following ratios of parameters:
\begin{align} 
&\frac{a^c_1}{a^c_2}=-0.20^{+3.93}_{-0.98}\,,\quad \frac{a^c_3}{a^c_2}=3.12^{+3.78}_{-15.05}\,,\quad \chi^2_{\text{d.o.f}}=0.13\,,\label{fchc}\\
&\frac{a^b_1}{a^b_2}=0.26^{+5.29}_{-2.30}\,,\quad \frac{a^b_3}{a^b_2}=1.84^{+8.48}_{-20.43}\,,\quad \chi^2_{\text{d.o.f}}=0.14\,,\label{fchb}
\end{align}
where the superindex $c$ and $b$ label the results for the $\psi(2S)\to J/\Psi \pi^+\pi^-$ and $\Upsilon(2S)\to \Upsilon(1S) \pi^+\pi^-$ transitions respectively. These fits are performed using the relativistic kinematics [see \eqs{rrep1}{rrep2}]. If instead we use nonrelativistic kinematics (see \eqs{rrep3}{rrep4}), fits of very similar quality are obtained. The results are shown in Fig.~\ref{cheft} together with the experimental data. Note the small $\chi_{\text{d.o.f}}^2$ obtained in the fits to both transitions. The uncertainties quoted in \eqs{fchc}{fchb} correspond to the range of parameter values with $\chi^2_{\text{d.o.f}}\leq 1$, which is shown in Fig.~\ref{corrplot}. In this figure, it can be appreciated that there is a strong correlation between $a^Q_1/a^Q_2$ and $a^Q_3/a^Q_2$. Therefore, it would be wrong to consider the error of $a^Q_1/a^Q_2$ and $a^Q_3/a^Q_2$ independently. Instead, the allowed region of parameter space is given by the dashed region in Fig.~\ref{corrplot}. This corresponds to a particular linear combination of $a^Q_1/a^Q_2$ and $a^Q_3/a^Q_2$. We conclude then that, with the present data on the width spectrum, we cannot simultaneously fit both parameter ratios with high accuracy . 

If we compare these fits with the ones in Ref.~\cite{Mannel:1995jt}, we observe some differences. The authors in Ref.~\cite{Mannel:1995jt} report that the best fit yields $a_3=0$  ($g_3$ in their notation). This does not coincide with our best fit, but is consistent with the uncertainty. The ratio $a_1/a_2$ corresponds to $1+g_1/(2g)$ in the notation of Ref.~\cite{Mannel:1995jt} and their best fits correspond to $a^c_1/a^c_2\sim 1.2$ and $a^b_1/a^b_2\sim 1.1$. This is compatible with our fits considering the uncertainty. One should keep in mind that the experimental data source in Ref.~\cite{Mannel:1995jt} is Ref.~\cite{Albrecht:1986gb}, whereas we use more recent data~\cite{Alexander:1998dq}\footnote{Very similar results to those in Ref.~\cite{Mannel:1995jt} were obtained in Ref.~\cite{Guo:2004dt} using more recent experimental data \cite{Bai:1999mj,Alexander:1998dq} including pion final state interactions through unitarization of $\chi$PT.}.

We now use the amplitude in \eq{qumqum2p} to obtain the normalized differential decay width and fit it to the experimental data. Equation~\eqref{qumqum2p} is equivalent to \eq{2pgp1} with the parameters $a_i$ taken as in Eqs.~(\ref{c21a1})-(\ref{c21a3}). The normalized decay width is independent of $\beta^{(12)}_{r,Q}$, and its functional form depends only on the parameter $\kappa$. Thus, the hadronic pNRQCD, which incorporates the multipole expansion, and a dimensional counting for the overlap of the hybrids with gluonic operators, is more predictive than the EFT relying on chiral and spin symmetry only, since the number of free parameters is reduced from two to one. Note that our approach yields the same normalized decay width line shape, as the one obtained using the twist expansion.

We fit the line shapes of the charmonium and bottomoniun data independently. Since the value of $\kappa$ should be independent of the heavy quarkonium dynamics, we also perform a simultaneous fit to both data sets. Using the nonrelativistic kinematics, we obtain the following values for $\kappa$:

\begin{align}
&\kappa_c=0.243^{+0.014}_{-0.013} \,,\quad \chi^2_{\text{d.o.f}}=0.18\,,\\
&\kappa_b=0.219^{+0.016}_{-0.015}\,,\quad \chi^2_{\text{d.o.f}}=0.14\,,\\
\label{kappajointnorel}
&\kappa_{\rm joint}=0.233\pm 0.013\,,\quad \chi^2_{\text{d.o.f}}=0.18\,.
\end{align}

The range of values correspond to the range with $\chi^2-\chi_{\rm min}^2\leq 1$. If we use the relativistic kinematics, the fits yield slightly different values,

\begin{align}
&\kappa_c=0.277\pm 0.015\,,\quad \chi^2_{\text{d.o.f}}=0.17\,,\\
&\kappa_b=0.229\pm 0.016 \,,\quad \chi^2_{\text{d.o.f}}=0.14\,,\\
\label{kappajointrel}
&\kappa_{\rm joint}=0.247^{+0.014}_{-0.013}\,,\quad \chi^2_{\text{d.o.f}}=0.25\,.
\end{align}

\begin{figure}[ht!]
\includegraphics[width=.6\textwidth]{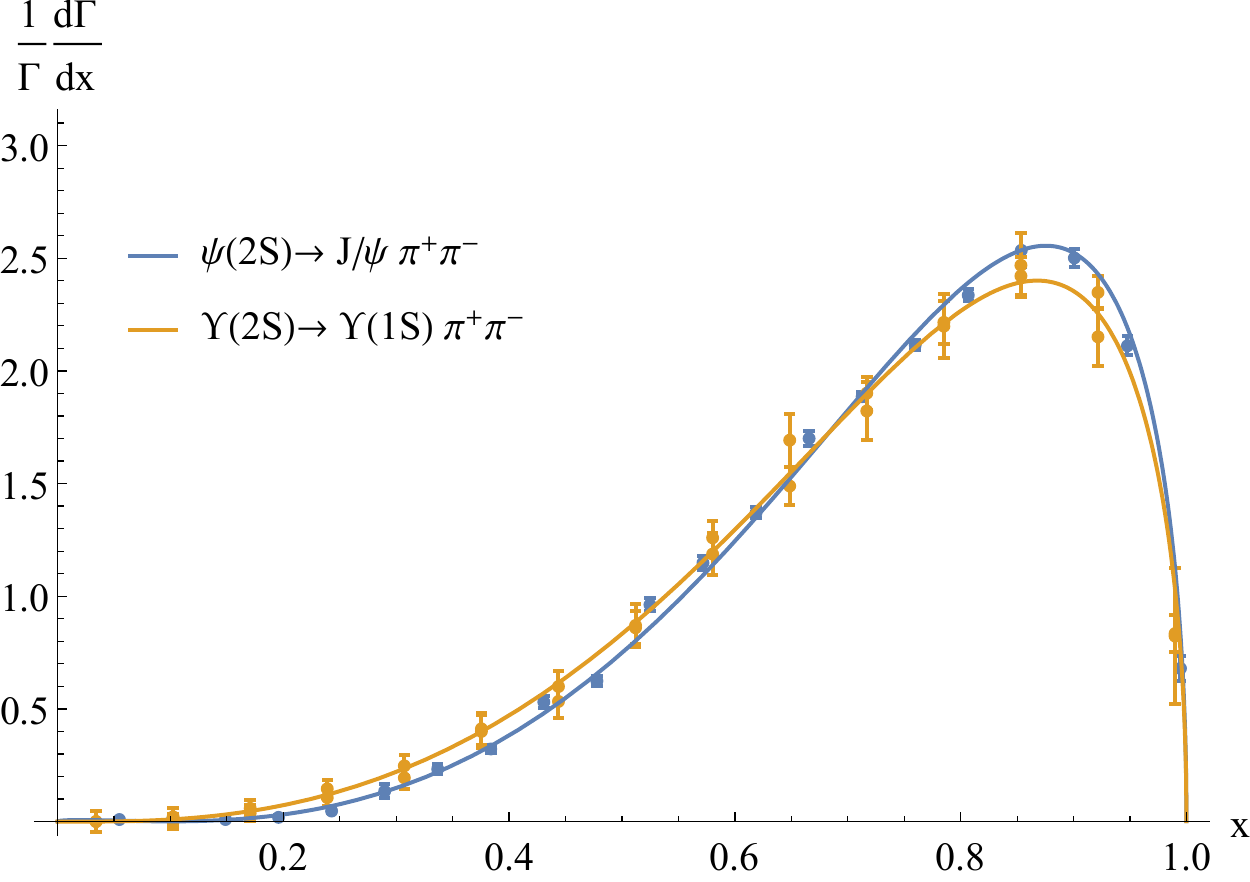}\\
\caption{Plot of the normalized differential decay width spectrum. The dots are the experimental data for $\psi(2S)\to J/\Psi \pi^+\pi^-$~\cite{Aaboud:2016vzw} and $\Upsilon(2S)\to \Upsilon(1S) \pi^+\pi^-$~\cite{Alexander:1998dq} in blue and yellow, respectively. In the same color scheme, the continuous lines are the fits of the theoretical expression obtained from the amplitude in \eq{qumqum2p} computed with the hadronic pNRQCD Lagrangian. The variable $x$ is defined as $x=\frac{m_{\pi\pi}-2m_{\pi}}{m_{2S}-m_{1S}-2m_{\pi}}$.}
\label{fig:kappa}
\end{figure}

The differences with the nonrelativistic fit are of the order of the difference between the charmonium and bottomonium fit. This is reasonable, as they both measure relativistic effects. We will take the difference between Eqs.~\eqref{kappajointnorel} and \eqref{kappajointrel} as a measure of the size of subleading effects. We then combine it in quadrature with the statistical error quoted in \eq{kappajointrel} and give
\be
\label{kappafinal}
\kappa=0.247(20)
\ee
as our default value. In any case, it is remarkable that all fits yield similar values for $\kappa$, which we take as a confirmation of the independence of $\kappa$ of the heavy quarkonium dynamics.  For illustration, we show the plots of the fit using relativistic kinematics for charmonium and bottomonium data in Fig. \ref{fig:kappa}. Actually,  we can also observe  in Figs.~\ref{cheft} and \ref{fig:kappa} the similarity of the experimental data for the line shapes of the charmonium and bottomonium spectra. This can be taken as a reflection of the independence of this observable on the heavy quarkonium dynamics, which is a prediction of the effective theory. 

Previous fits of the decay width spectrum of $\psi(2S)\to J/\psi\pi\pi$ have been carried out in Ref.~\cite{Bai:1999mj} using the transition amplitude from Ref.~\cite{Novikov:1980fa}, and in Ref.~\cite{Chen:2019hmz} in which the pion final state interactions were taken into consideration through an Omn\`es function. The reported values are $\kappa_c=0.186(3)$ and $\kappa_c=0.135(5)$, respectively. We have checked that the main source of this discrepancy is that the transition amplitudes used in those references do not include the complete ${\cal O}(p^2)$ pion mass contribution: in Ref.~\cite{Chen:2019hmz}, the coefficient $a_3$ is set to zero, and in Ref.~\cite{Bai:1999mj}, terms proportional to the pion mass have been set to zero. 

\subsection{Total decay widths}
\label{Sec:Total}
The expressions for the total decay width in terms of the chiral coefficients $a_i$ can be found in \eqs{lnshp}{lnshpint}. Inserting the values from \eq{fchc} or \eq{fchb}, the  remaining free parameter, $a_2$, can be adjusted to reproduce the experimental value of the total decay width. In fact, since the total width is a quadratic function in $a_2$, two solutions are possible,
\begin{align}
&a^c_2=-25.88\;{\rm GeV}^{-3}\,,\quad a^c_2=25.52\;{\rm GeV}^{-3}\,,\\
&a^b_2=-10.98\;{\rm GeV}^{-3}\,,\quad a^b_2=10.78\;{\rm GeV}^{-3}\,.
\end{align}
\indent
In our EFT, the total decay width is proportional to $(\beta^{(12)}_{r,Q})^2$ (note that it is at this level where there is a difference with predictions using the twist expansion). This object is dependent on the wave functions of heavy quarkonium and hybrids, as well as on the energy difference among them. Typically, this quantity will suffer from rather large uncertainties, as we are forced to make strong approximations to compute this object. We neglect the effect of higher gluelump excitations with the same quantum numbers. We expect that higher gluelumps will give smaller contributions, as they are suppressed by larger energy differences. Still, this is an approximation. In some circumstances, it can be compulsory to sum all of them to recover some high energy logarithms. Nevertheless, compared with other uncertainties this effect will be small. Therefore, in this paper, we neglect the error associated to neglecting higher gluelump channels. We account for the error associated to the energy splitting between singlet and hybrid states using the error of $\Lambda_1$, the lowest gluelump mass (for further details, see Appendix \ref{AppStates}). For the lowest hybrid, we compute $\beta^{(12)}_r$ with the $m^3\mathcal{S}_1$ hybrid intermediate states $m=1,...,4$ (we observe that the effect of introducing three or four hybrid states is comparatively small compared with other uncertainties). For the static potential of the singlet and hybrid, we take two possible parametrization that we explain in further detail in Appendix \ref{Sec:anomaly}. One aims for a good description  at all distances of the lattice-evaluated static singlet and hybrid potentials, but constrained to have the right behavior at short distance. The other parametrization keeps the shape of the potentials predicted by the multipole expansion in the whole fitted range. Throughout the paper, we will take the former for our central values and the difference with the latter as an estimate of the error associated to the approximate knowledge of the singlet and hybrid wave functions and binding energies. We obtain the following values for $\beta^{(12)}_{r,Q}$:
\begin{align}
&\beta^{(12)}_{r,c}=11.46(_{+1.81}^{-1.36})_{\Lambda_1}(\pm 3.03)_{\rm s.p.}\,,{\rm GeV}^{-3}\,,\\
&\beta^{(12)}_{r,b}=3.45(_{+0.43}^{-0.34})_{\Lambda_1}(\pm 0.97)_{\rm s.p.}\,{\rm GeV}^{-3}\,.
\end{align}
The uncertainties are labeled according to the source: $\Lambda_1$ (the lowest lying gluelump mass), and s.p. (the different parametrization for the singlet and hybrid static potentials). At this point, it is worth mentioning that $\beta^{(11)}_{r,Q}$, relevant for heavy quarkonium-nucleon interaction and the chiral-associated heavy quarkonium energy shift, has also been analyzed using hybrid potentials in Ref. \cite{Lakhina:2003pj}.

Introducing these values of $\beta^{(12)}_{r,Q}$, and $\kappa=0.247(20)$, in Eqs.~\eqref{c21a1}-\eqref{c21a3}, the expressions for the total width in Eqs.~\eqref{lnshp} and \eqref{lnshpint} read (the experimental values are taken from \cite{Tanabashi:2018oca})
\begin{align}
\label{Gammacpipi}
&\Gamma_{\psi(2S)\to J/\psi\pi^+\pi^-}=46.2(_{+15.7}^{-10.3})_{\Lambda_1}(_{+3.4}^{-3.2})_{\kappa}(\pm 21.2)_{\rm s.p.}\,{\rm keV}\,, \quad \Gamma^{\rm exp}=102.1(2.9) \,{\rm keV}\,,\\
\label{Gammabpipi}
&\Gamma_{\Upsilon(2S)\to\Upsilon(1S)\pi^+\pi^-}=3.08(_{+0.81}^{-0.58})_{\Lambda_1}(_{+0.22}^{-0.23})_{\kappa}(\pm 1.49)_{\rm s.p.}\,{\rm keV}\,, \quad \Gamma^{\rm exp}=5.71(48) \,{\rm keV}\,.
\end{align}

Our numbers differ from the experimental ones by about a factor 2. One should keep in mind however that our estimates suffer from large uncertainties. We find a significant dependence on variations of the wave function of the hybrid and singlet states. The error generated by the uncertainty on the energy difference between singlet and hybrid states is somewhat smaller.These error estimates are of the right magnitude, though not large enough, to completely account for the difference with experiment. One should keep in mind, however, that besides those errors already estimated, one error that has not been incorporated in this analysis is due to the uncertainties associated to the hadronization of the local operator: on the one hand, we have ${\cal O}(\alpha_s)$ corrections to \eqs{eeh2p}{bbh2p} due to the beta function. These ${\cal O}(\alpha_s)$ corrections are generated at a low-energy scale, which makes their evaluation not feasible. On the other hand, these effects factor out and could be reabsorbed in $(\beta^{(12)}_{r,Q})^2$ if we let this object to be a free parameter, not fixed by theory. Alternatively, these effects are independent of the heavy quarkonium dynamics and would cancel in the ratio $\Gamma_{\psi(2S)\to J/\psi\pi^+\pi^-}/\Gamma_{\Upsilon(2S)\to\Upsilon(1S)\pi^+\pi^-}$. We will discuss this ratio later in \Sec{Sec:Ratios}. Other corrections to \eqs{eeh2p}{bbh2p} are due to ${\cal O}(p^4)$ chiral corrections. We expect those not to be very important due to the limited phase space available. Other source of error comes from neglecting the anomalous dimension of the light-quark mass. These two sources of error would affect the determination of the line shapes. As we have obtained a pretty good fit for them, we will neglect these sources of error in the following.   

The other error we have not incorporated in this analysis, nor in those we will perform later, is the error due to the working hypothesis we use in this paper (saturation of the interpolating field by those with smaller dimensionality). The reason is that  we want to see whether such hypothesis is feasible, and if so what is the expected error, by comparing our predictions with experiment. 

\section{\texorpdfstring{$\bar Q Q(2S) \rightarrow \bar Q Q(1P) \pi$}{QbarQ(2)S->QbarQ(1S)pi} hadronic transitions}\label{2s1ptran}

The $\bar Q Q(2S) \to \bar Q Q(1P) \pi$ hadronic transition\footnote{Let us note that this type of transition was first considered in Ref.~\cite{Kuang:1981se} for the $\Upsilon(2S)\to h_b(1P)\pi^0$ decay. However, this particular case turned out not to be kinematically allowed.} is zero with the LO Lagrangian in \eq{pnrqcd1} considered this far. The first nonzero contribution is generated by spin-isospin breaking effects. The leading spin-dependent operators originate from the following pNRQCD operators:
\begin{align}
\delta L_{\rm pNRQCD} =& \int d^3R d^3r \,\left( \frac{gc_F}{2m_Q}\right){\rm Tr}\left[\left\{\S^{\dag},\,\bm{\sigma}\right\}\cdot\bm{B}\Oc+\Oc^{\dag}\bm{B}\cdot\left\{\bm{\sigma},\,\S\right\} \right]  
\,.
\label{pnrqcd2}
\end{align}
We remind that the singlet, octet, and hybrid fields in the Lagrangians that appear in this paper are organized in $SU(2)$ spin multiplets. An alternative representation of \eq{pnrqcd2} in terms of the spins of the quark and antiquark can be found in Eq. (105) of \cite{Brambilla:2002nu}. This last representation is the one we will customarily use for the computation of the matrix elements. 

The renormalization group improved expression for the matching coefficient $c_F$ is known with next-to-leading logarithmic accuracy \cite{Amoros:1997rx,Czarnecki:1997dz}. In order to include the leading isospin violation effects, we should no longer consider the light-quark masses degenerate in \eq{pnrqcd1}: we take $m_u=2.118$ MeV and $m_d=4.690$  MeV instead~\cite{Tanabashi:2018oca}.

Adding the operator in \eq{pnrqcd2} to the leading pNRQCD Lagrangian in \eq{pnrqcd1} and considering isospin violation effects produces the following new terms in the hadronic Lagrangian at LO\footnote{There is another possible pseudoscalar operator of the same order as $\langle\chi_-\rangle$: $\langle D_{\mu}u^{\mu}\rangle$. However they are both related through the leading order equations of motion $D_{\mu}u^{\mu}=i\left(\chi_--\langle\chi_-\rangle/2\right)/2$.}: 

\begin{align}
\delta L_{pNRQCD}^{had} = \int d^3Rd^3r \,{\rm Tr}\Bigl[&\left(\bm{r}\cdot \hat{\bm{r}}_{\la}S^{\dag}\Psi_{1^{+-}\la}+\text{h.c}\right)t^{(r1^{+-})}iF\langle\chi_-\rangle\nn\\
&+\left(S^{\dag}\left\{\bm{\sigma}\cdot \hat{\bm{r}}_{\la},\Psi_{1^{+-}\la}\right\}+\text{h.c}\right)t^{(S1^{+-})}\Bigr]\,.
\label{deltaLpwave}
\end{align}
Terms that break spin and isospin symmetry simultaneously are not considered, since they produce subleading contributions to the transition we are considering.

Let us match the parton and hadronic description of the spin-dependent mixing operator. In the hadronic EFT, the correlator reads
\begin{align}
\langle 0|\Psi_{1^{+-}\la}(t,\,\bm{r},\,\bm{R})S^{\dag}(0,\,\bm{r},\,\bm{R})|0 \rangle_{\rm amp.}=i \left(\bm{\sigma}_1-\bm{\sigma}_2\right)\cdot \hat{\bm{r}}^{\dag}_{\la}t^{(S1^{+-})}\,,\label{sdmoa}
\end{align}
and in the partonic version of pNRQCD, we have
\begin{align}
&\langle 0|\hat{\bm{r}}^{\dag}_{\la}\cdot G_{1^{+-}}^{a\dag}(t,\bm{R})O^{a}\left(t,\,\bm{r},\,\bm{R}\right)S^{\dag}(0,\,\bm{r},\,\bm{R})|0\rangle_{\rm amp.}\nn\\
&=i\frac{g c_F}{2m_Q}\sqrt{\frac{T_F}{N_c}}\left(\bm{\sigma}_1-\bm{\sigma}_2\right)^i \hat{\bm{r}}^{j\dag}_{\la}\langle 0|G_{1^{+-}}^{ai\dag}(\bm{R})\bm{B}^{aj}(\bm{R})|0\rangle 
\,,\label{sdmob}
\end{align}
where \textit{amp.} signals that only amputated contributions are considered. Now we consider $G_{1^{+-}}^{a}$ to be given by a sum of all possible gluonic operators with the same quantum numbers
\begin{align}
G_{1^{+-}}^{a}=Z^{-1/2}_B g\bm{B}^a+Z^{-1/2}_{D\times E}\left(\bm{D}\times g\bm{E}\right)^a+\cdots\;.
\end{align}
Similarly, to the previous section we hypothesize that there is a correlation between the dimensionality of the interpolating operator and the strength of the interpolation with the hybrid, such that higher dimension operators are subleading, so the series can be truncated at LO. Using this truncation and \eq{glmpn} in \eq{sdmob}, and matching to \eq{sdmoa}, we arrive at
\begin{align}
t^{(S1^{+-})}=\frac{ c_F}{2m_Q}\sqrt{\frac{T_F Z_B}{N_c}}\,.
\end{align}

Let us now match the operators with an odd number of pions in the Lagrangian in \eq{deltaLpwave}. In the hadronic theory, we have
\begin{align}
\int d^4 x e^{i p\cdot x}\langle US|\pi^0(x)S(t,\,\bm{r},\,\bm{R})\Psi^{\dag}_{1^{+-}\la}(0,\,\bm{r},\,\bm{R})|US\rangle_{\rm amp.}=-it^{(r1^{+-})}\bm{r}\cdot \hat{\bm{r}}_{\la}\,2 m^2_{\pi}\frac{m_d-m_u}{m_u+m_d}\,,\label{1psdma}
\end{align}
and in pNRQCD,
\begin{align}
&\langle \pi^0(p)|S(t,\,\bm{r},\,\bm{R})\hat{\bm{r}}_{\la}\cdot G^a_{1^{+-}}(0,\,\bm{R}) O^{a\dag}(0,\,\bm{r},\,\bm{R})|0\rangle_{\rm amp.}=ig\sqrt{\frac{T_F}{N_c}}\hat{\bm{r}}^i_{\la}\bm{r}^j\langle \pi^0|\bm{E}^{aj}(\bm{R})G^{ai}_{1^{+-}}(\bm{R})|0\rangle\nn\\
&=\frac{i}{3}\sqrt{\frac{T_F}{N_cZ_B}}\hat{\bm{r}}_{\la}\cdot\bm{r}\langle \pi^0|g^2\bm{E}\cdot\bm{B}|0\rangle=\frac{i}{3}4\pi^2\frac{m_d-m_u}{m_d+m_u}F m^2_{\pi}\sqrt{\frac{T_F}{N_cZ_B}}\hat{\bm{r}}_{\la}\cdot\bm{r}\,,\label{1psdmb}
\end{align}
where in the last step we have made use of the results of Appendix~\ref{Sec:AxialAnomaly} for the hadronization of the gluonic operator through the axial anomaly. Comparing Eqs.~\eqref{1psdma} and \eqref{1psdmb}, we arrive at
\begin{align}
t^{(r1^{+-})}=-\frac{2\pi^2}{3} F
\sqrt{\frac{T_F}{N_cZ_B}}\,.
\end{align}

\begin{figure}[ht!]
   \centerline{\includegraphics[width=.8\textwidth]{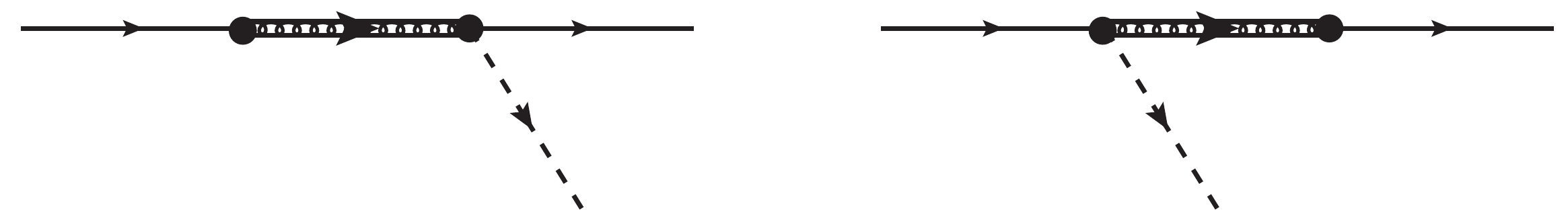}}
	\caption{Single and dashed lines represent quarkonia and pions. The double line with a curly line inside represents hybrid states.}
	\label{optf}
 \end{figure}

We are interested in investigating the decay $\psi(2S)\to h_c(1P)\pi^0$, which is given in the hadronic pNRQCD EFT by the diagrams in Fig.~\ref{optf}. The matrix element reads

\begin{align}
i\mathcal{A}&=-F m^2_{\pi}\frac{2\pi^2c_F}{3m_Q}\frac{T_F}{N_c}\frac{m_d-m_u}{m_u+m_d}\beta^{(12)}_{\sigma}\,,\label{1ptqm}
\end{align}
with
\begin{align}
\beta^{(12)}_{\sigma}\equiv \sum_{m}&\left(\langle h_c(1^1P_1)|\hat{\bm{r}}_{\la}\cdot \bm{r} |\Psi_m\rangle\frac{i}{m_{h_c}-m_m}\langle\Psi_m| \hat{\bm{r}}^{\dag}_{\la}\cdot\left(\bm{\sigma}_1-\bm{\sigma}_2\right)|\psi(2^3S_1)\rangle
\right.\nn \\
&\left.+\langle h_c(1^1P_1)|\hat{\bm{r}}_{\la}\cdot\left(\bm{\sigma}_1-\bm{\sigma}_2\right) |\Psi_m\rangle\frac{i}{m_{\psi(2S)}-m_m}\langle\Psi_m|\hat{\bm{r}}^{\dagger}_{\la}\cdot\bm{r}|\psi(2^3S_1)\rangle\right)\,.
\end{align}
In the first term, the $\psi(2S)$ mixes into a hybrid state with $J^{PC}=1^{--}$ but in a spin singlet, then the hybrid decays into a $h_c$ conserving the spin but changing the angular momentum. The intermediate hybrid states fulfilling these conditions are $m^1\mathcal{P}_1$. This term turns out to be small because the second vertex only has contributions from the $\la=0$ piece of the hybrid state, which is small for $^1\mathcal{P}_1$ states. In the second term, the order is switched; in this case the allowed hybrid intermediate states are $m^3\mathcal{S}_1$. Note that all the hybrid states appearing as intermediate states in this transition are associated to the $1^{+-}$ gluelump. As in the two-pion decays, the normalization factors $Z_B$ cancels out. The matrix elements are given in Appendix~\ref{app:matelem}. 

We compute the amplitude for up to $m=4$ for the ${}^3\mathcal{S}_1$ and up to $m=2$ for the $^1\mathcal{P}_1$ hybrid intermediate states. We use the renormalization group improved expression of $c_F(1~{\rm GeV})$=1.12155 up to next-to-leading logarithmic order, where we have used the values $\alpha_s(1~{\rm GeV})$ = 0.4798 and $\alpha_s(1.496~{\rm GeV})$ = 0.3522. The decay width can be obtained using the amplitude of \eq{1ptqm} in \eq{dw1p},
\begin{align}
\label{eq:hc}
\Gamma_{\psi(2S)\to h_c(1P)\pi^0}=104(_{+80}^{-35})_{\rm \Lambda_1}(\pm 21)_{\rm l.q.}(\pm 1)_{\rm s.p}\,{\rm eV}\,,\quad \Gamma^{\rm exp}=255(39)\,{\rm eV}\,.
\end{align}
The error analysis has been performed similarly to the previous section. In this case, we do not have error associated to $\kappa$, but still have a dependence on $\Lambda_1$ (the lowest lying gluelump mass),  and s.p. (the different parametrization for the singlet and hybrid static potentials). We also estimate the error due to the value of the light-quark (l.q.) mass ratio, which we take as
\begin{align}
\frac{m_d-m_u}{m_d+m_u}=0.35\pm 0.07\,,\label{lqmu}
\end{align}
with the value of $m_d/m_u$ from the PDG.

This observable is interesting. Unlike previous decays, it does not suffer from the uncertainties associated to the hadronization of the local operator: The axial anomaly does not get ${\cal O}(\alpha)$ corrections, nor there are ${\cal O}(p^4)$ chiral corrections. We also observe a very weak dependence on variations of the wave function of the hybrid and singlet states. The major error is generated by the uncertainty on the energy difference between singlet and hybrid states. Once this error is taken into account, the result is roughly compatible with experiment. 

In principle, for $\bar Q Q(2S) \rightarrow \bar Q Q(1P) \pi$ transitions, the energy release is small so one can use the twist expansion with no fear if one assumes that $\lQ \ll mv^2$. We then would have an alternative determination with which one can compare. In this respect, let us note that the only previous theoretical estimate was $\Gamma_{\psi(2S)\to h_c(1P)\pi^0}\sim 15$ eV, from Ref.~\cite{Voloshin:2007dx}, using the twist expansion ($\lQ \ll mv^2$ and $E \ll mv^2$). 

\section{\texorpdfstring{$\bar Q Q(2S) \rightarrow \bar Q Q(1S) \pi^0(\eta)$}{QbarQ(2S)->QbarQ(1S)pi(eta)} hadronic transitions}
\label{2s1s1pitran} 

We now turn our attention to $\bar Q Q(2S) \rightarrow \bar Q Q(1S) \pi^0(\eta)$ hadronic transitions. Since these transitions break spin symmetry, they are zero at LO in the EFT. The leading contribution to these decays is generated by
\begin{align}
\delta L_{\rm pNRQCD} =& \int d^3R\, d^3r \,\frac{gc_F}{4m_Q}{\rm Tr}\left[\left[\S^{\dag}\,,\bm{\sigma}\right]\cdot\left(\bm{r}^l\bm{D}_l\bm{B}\right)\Oc+\Oc^{\dag}\left(\bm{r}^l\bm{D}_l\bm{B}\right)\cdot\left[\bm{\sigma},\S\right]\right]
\,,
\label{pnrqcd3}
\end{align}
which has to be added to \eq{pnrqcd1}. In the hadronic EFT, these operators correspond to
\begin{align}
\delta L_{pNRQCD}^{\rm had} &= \int d^3Rd^3r \,{\rm Tr}\left[
S^{\dagger}\left[\bm{\sigma}^i,\,\Psi_ {1^{--}\la}\right]+\text{h.c}\right]
\nn
\\
&
\times
\left(t^{(da1^{--})}\hat{\bm{r}}^i_\la\bm{r}^j+t^{(db1^{--})}\hat{\bm{r}}^j_\la\bm{r}^i+t^{(dc1^{--})}\hat{\bm{r}}_\la\cdot\bm{r}\delta^{ij}\right)iF\bm{\partial}^j\langle\chi_-\rangle\,.
\label{LhadPwave}
\end{align}

To determine the Wilson coefficients, we compute the transition amplitude both in the hadronic and partonic versions of the effective theory. Nevertheless, there is one extra subtlety to be taken into account. The $\pi^0$ and $\eta$ fields in \eq{pions} mix, and to obtain the physical states the mass matrix needs to be diagonalized. The physical states correspond to
\begin{align}
&\pi^0_{\rm phys.}=\pi^0+\epsilon\eta\,,\\
&\eta_{\rm phys.}=\eta-\epsilon\pi^0\,,
\end{align}
with $\epsilon$ the mixing angle
\begin{align}
\epsilon=\frac{\sqrt{3}(m_d-m_u)}{4m_s+2(m_u+m_d)}\,.
\end{align}
To compute the decay amplitudes, the physical states must be considered. In the transitions with $\eta$ emission, the contribution due to the mixing is subleading, but in the case of $\pi^0$ emission both contributions are of the same order. In practice, this amounts to an extra factor $3/2$ in the $\pi^0$ decay amplitude. In the hadronic EFT, we have
\begin{align}
&\int d^4xe^{ip\cdot x}\langle 0|\pi^0(x)\Psi_{1^{--}\la}(t,\,\bm{r},\,\bm{R})S^{\dag}(0,\,\bm{r},\,\bm{R})|0\rangle_{\rm amp.}=-3i\frac{m_d-m_u}{m_u+m_d}m^2_{\pi}\nn\\
&\times(\bm{\sigma}_1+\bm{\sigma}_2)\cdot\left[t^{(da1^{--})}\hat{\bm{r}}_\la (\bm{r}\cdot\bm{p}_{\pi})+t^{(db1^{--})}\bm{r}(\hat{\bm{r}}_\la \cdot\bm{p}_{\pi})+t^{(dc1^{--})}\bm{p}_{\pi} (\bm{r}\cdot\hat{\bm{r}}_\la)\right]\,,\label{pwa1}
\end{align}
and in the partonic pNRQCD, 
\begin{align}
&\langle\pi^0(p)|\hat{\bm{r}}^\dag_{\la}\cdot G^{a\dag}_{1^{--}}(t,\bm{R})O^{a}(t,\,\bm{r},\,\bm{R})S^{\dag}(0,\,\bm{r},\,\bm{R})|0\rangle_{\rm amp.}\nn\\
&=\sqrt{\frac{T_F}{N_c}}\frac{igc_F}{4m_Q}\bm{r}^j(\bm{\sigma}_1+\bm{\sigma}_2)^k\langle\pi^0(p_{\pi})|\hat{\bm{r}}^\dag_{\la}\cdot G^{a\dag}_{1^{--}}(\bm{D}_j\bm{B}_k)^a|0\rangle\nn\\
&=\sqrt{\frac{T_F}{Z_EN_c}}\frac{c_F}{m_Q}\frac{2\pi^2}{15}\frac{m_d-m_u}{m_u+m_d}m^2_{\pi}F\left[3(\bm{\sigma}_1+\bm{\sigma}_2)\cdot\hat{\bm{r}}_\la \bm{r}\cdot\bm{p}-(\bm{\sigma}_1+\bm{\sigma}_2)\cdot\bm{r}\hat{\bm{r}}_\la \cdot\bm{p}\right]\,,\label{pwa2}
\end{align}
where in the last step we have used \eq{1mmgle} truncated to the first term and the anomaly relation in Appendix \ref{Sec:PwaveAnomaly} [note that this relation does not suffer from ${\cal O}(\alpha_s)$ corrections]. Matching Eqs.~\eqref{pwa1} and \eqref{pwa2}, we obtain
\begin{align}
&t^{(da1^{--})}= -\sqrt{\frac{T_F}{Z_EN_c}}\frac{c_F}{m_Q}\frac{2\pi^2}{15}F\,,\\
&t^{(db1^{--})}= \sqrt{\frac{T_F}{Z_EN_c}}\frac{c_F}{m_Q}\frac{2\pi^2}{45}F\,,\\
&t^{(dc1^{--})}=0\,.
\end{align}

We are interested in investigating the decays $\psi(2S)\to J/\psi\pi^0$, $\Upsilon(2S)\to \Upsilon(1S)\pi^0$, as well as the decays $\psi(2S)\to J/\psi\eta$, $\Upsilon(2S)\to \Upsilon(1S)\eta$. The diagrams involved in the decay are drawn in Fig.~\ref{optf}. These contain the $t^{(r 1^{--})}$ vertex from \eq{bolag2} mixing the quarkonia into a hybrid associated to the $1^{--}$ gluelump, and the $P$-wave pion emission vertices from \eq{LhadPwave}. The former vertex conserves the spin state and is only nonvanishing for $l=\ell$. Therefore, the intermediate hybrid states for the transitions we are considering must be $^3\mathcal{S}_1$. The amplitude for a pion emission then reads
\begin{align}
i\mathcal{A}=-i\frac{T_F}{N_c}\frac{c_F}{m_Q}\frac{8\pi^2}{45}\frac{m_d-m_u}{m_d+m_u}F_\pi m^2_{\pi}\left(\epsilon^{*}_{1S}\times\epsilon_{2S}\right)\cdot\bm{p}\beta^{(12)}_{r,Q}\,.\label{pwpia}
\end{align}
Note that the factor involving the sum over intermediate hybrid states: $\beta^{(12)}_{r,Q}$, given in \eq{betarc21}, is the same to the one appearing in the two-pion transitions. $\epsilon_{mS}$ stands for the polarization of a $m^3S_1$ quarkonium state. The decays amplitude to one $\eta$ is
\begin{align}
i\mathcal{A}=-i\frac{T_F}{N_c}\frac{c_F}{m_Q}\frac{8\pi^2}{45\sqrt{3}}\left(m^2_{\eta}-m^2_{\pi}\right)F\left(\epsilon^{*}_{1S}\times\epsilon_{2S}\right)\cdot\bm{p}\beta^{(21)}_{r,Q}\,.\label{pwetaa}
\end{align}

We have computed the decay widths by summing over final polarizations and averaging over the incoming ones the square of the amplitudes in Eqs.~\eqref{pwpia} and \eqref{pwetaa}, and inserting the result in \eq{dw1p}. We use relativistic kinematics for the outgoing quarkonium state in the phase space calculation and the renormalization group improved expression of $c_F$ with next-to-leading logarithmic accuracy. The expression for charmonium can be found in the previous section. For bottomonium, we have $c_F(1~{\rm GeV})$=0.87897 using $\alpha_s(4.885~{\rm GeV})$ = 0.2148. $\beta^{(21)}_{r,Q}$ is computed considering up to four hybrid intermediate states. The results obtained for the decays are the following:
\begin{align}
\label{Gamma2Spi1}
&\Gamma_{\psi(2S)\to J/\psi \pi^0}=40(_{+14}^{-9})_{\Lambda_1}(\pm 8)_{\rm l.q.}(\pm 18)_{\rm s.p.}\,{\rm eV}\,,& &\Gamma^{\rm exp}=373(14)\,{\rm eV}\,,\\
&\Gamma_{\psi(2S)\to J/\psi \eta}=1.19(_{+0.41}^{-0.27})_{\Lambda_1}(\pm 0.5)_{\rm s.p.}\,{\rm keV}\,,& &\Gamma^{\rm exp}=9.91(30)\,{\rm keV}\,,\\
&\Gamma_{\Upsilon(2S)\to \Upsilon(1S) \pi^0}=0.21(_{+0.06}^{-0.04})_{\Lambda_1}(\pm 0.05)_{\rm l.q.}(\pm 0.10)_{\rm s.p.}\,{\rm eV}\,,& &\Gamma^{\rm exp}<1.28\,{\rm eV}\,,\\
\label{Gamma2Seta2}
&\Gamma_{\Upsilon(2S)\to \Upsilon(1S) \eta}=1.58(_{+0.42}^{-0.80})_{\Lambda_1}(\pm 0.76)_{\rm s.p.}\,{\rm eV}\,,& &\Gamma^{\rm exp}=9.3(1.5)\,{\rm eV}\,.
\end{align}
The subindices label the source of the uncertainty. The error analysis is equal to the one performed in the previous section. Again we do not have error associated to $\kappa$, but still have a dependence on $\Lambda_1$ (the lowest lying gluelump mass),  and s.p. (the different parametrization for the singlet and hybrid static potentials). We also estimate the error due to the value of the light-quark mass ratio. 

As in the previous section, these decays do not suffer from the uncertainties associated to the hadronization of the local operator: the axial anomaly does not get ${\cal O}(\alpha_s)$ corrections, nor there are ${\cal O}(p^4)$ chiral corrections. On the other hand, unlike in the previous section, we find a significant dependence on variations of the wave function of the hybrid and singlet states. Indeed, the dependence is the same as the one we had in Sec. \ref{Sec:Total}. The error generated by the uncertainty on the energy difference between singlet and hybrid states is of the same order. Nevertheless, these error estimates are not large enough to account for the difference with experiment, particularly for the charmonium case. We later retake this issue when we consider ratios of decay rates. 

\section{Ratios}
\label{Sec:Ratios}

So far we have seen that the hadronic $S$- to $P$-wave heavy quarkonium transitions are roughly compatible with theory within errors. For decays to $S$-wave heavy quarkonium, we have seen that the overall magnitude of our predictions is smaller than experiment, specially for the $Q\bar Q(2S) \to Q\bar Q(1S)\pi^0(\eta)$ decays. Nevertheless, the normalization of the decays are the most uncertain object in our predictions. Therefore, we expect several uncertainties to cancel for ratios.  In fact, we have already seen in \Sec{Sec:Line-shape} that the normalized differential decay rates are well described by theory. Going further in this direction, we may try to explore the magnitude of the corrections associated to the different approximations we have made in this paper by studying different ratios. This we do in the following. 
  
The ratios of the $Q\bar Q(2S) \to Q\bar Q(1S)\pi^0(\eta)$ transitions do not depend on $\beta^{(12)}_{r,Q}$, and are completely determined at LO by chiral symmetry. Indeed the same result is obtained using a pure chiral approach or by using the twist expansion (though the use of the latter suffers from the same drawback as its use in $Q\bar Q(2S) \to Q\bar Q(1S)\pi\pi$ transitions \cite{Luty:1993xf}), as studied previously in Refs.~\cite{Ioffe:1981qa,Donoghue:1992ac,Casalbuoni:1992fd}. What changes is the overall coefficient, which cancels in the ratio. The only uncertainties affecting the ratios are the ones associated to the light-quark mass values, and possible chiral corrections to the amplitude affecting the ratio $F_{\eta}/F_{\pi}$ and the mixing angle $\epsilon$. Generically, we expect these to be of $\mathcal{O}((m_\pi,m_{K})/\Lambda_{\chi})\sim 14-50\%$. We can check this by comparing the theoretical and experimental ratios (for these and for the experimental ratios below we add the error quadratically) 
\begin{align}
\label{Rcchiral}
&R_{c\,\chi}\equiv R \left(\frac{\psi(2S)\to J/\psi \pi^0}{\psi(2S)\to J/\psi \eta}\right)=3.34(\pm 0.67)_{\rm l.q.}\times 10^{-2} \,,&&R_{c\,\chi}^{\rm exp}=3.76(18)\times 10^{-2}\,,\\
\label{Rbchiral}
&R_{b\,\chi}\equiv R\left(\frac{\Upsilon(2S)\to \Upsilon(1S) \pi^0}{\Upsilon(2S)\to \Upsilon(1S) \eta}\right)=13.5(\pm 2.7)_{\rm l.q.}\times 10^{-2} \,,&&R_{b\,\chi}^{\rm exp}<13.8(2.2)\times 10^{-2}\,.
\end{align}
Since the theoretical values are compatible experiment we conclude that chiral corrections are not needed at this level of precision.

It is also interesting to compare the theoretical and experimental ratios of the two-pion transitions of Sec.~\ref{ddws} over the one-pion or eta transition computed in Sec.~\ref{2s1s1pitran}, since this set of ratios is also independent of $\beta_{r,Q}^{(12)}$, or, in other words, of the heavy quarkonium dynamics (which is also the case with the twist expansion). The hadronization of the two-pion production using the energy-momentum tensor anomaly has corrections of $\mathcal{O}(\alpha_s)$ at a low-energy scale, as well as contributions of the anomalous dimension of the light-quark mass operator, both of which could be potentially large.\footnote{In principle there are also ${\cal O}(p^4)$ chiral corrections. Nevertheless, we expect those to be comparatively small due to the limited phase space available.} On the other hand the determination of the one single pion or eta matrix element through the axial anomaly only has chiral corrections, as discussed in the previous paragraph, and, in the case of one pion in the final states, the uncertainty on $m_u/m_d$. Hence, we explore whether comparing the ratios of the two- and one-pion (or eta) transitions to their experimental values allows us asses the size of the corrections to \eqs{eeh2p}{bbh2p}.
\begin{align}
\label{Rcpi}
R_{c,\pi}\equiv &
R\left(\frac{\psi(2S)\to J/\psi \pi^+\pi^-}{\psi(2S)\to J/\psi \pi^0}\right)=1161(\mp 232)_{\rm l.q.}(_{+84}^{-80})_{\kappa}\,,&& 
R_{c,\pi}^{\rm exp}=274(13)\,,\\
R_{c,\eta}\equiv&R\left(\frac{\psi(2S)\to J/\psi \pi^+\pi^-}{\psi(2S)\to J/\psi \eta}\right)=38.8(_{+2.8}^{-2.7}) _{\kappa} \,,&& 
R_{c,\eta}^{\rm exp}=10.3(4)\,,\\
R_{b,\pi}\equiv &R\left(\frac{\Upsilon(2S)\to \Upsilon(1S) \pi^+\pi^-}{\Upsilon(2S)\to \Upsilon(1S) \pi^0}\right)=14.4 (\mp 2.9)_{\rm l.q.}(^{-1.0}_{+1.1})_{\kappa} \times 10^3\,,&& 
R_{b,\pi}^{\rm exp}>4.5(4) \times 10^3\,,\\
\label{Rbeta}
R_{b,\eta}\equiv &R\left(\frac{\Upsilon(2S)\to \Upsilon(1S) \pi^+\pi^-}{\Upsilon(2S)\to \Upsilon(1S) \eta}\right)=1.94(^{-0.14}_{+0.15})_{\kappa} \times 10^3 \,,&& 
R_{b,\eta}^{\rm exp}=0.61(11) \times 10^3\,.
\end{align}
The differences with experiment are large, of the order of 75\%-60\% for the charmonium and bottomonium, respectively. In principle, we expect the most important uncertainties of these ratios to be due to the effect of higher order operators in the interpolating function of the hybrids and of the neglected ${\cal O}(\alpha_s)$ corrections in the hadronization of the local operators (this latter source of error only applies to the $Q\bar Q(2S) \to Q\bar Q(1S)\pi\pi$ transitions). In this respect, it is interesting to note that we expect the ${\cal O}(\alpha_s)$ corrections in the hadronization of the local operators to be largely independent of the bound state dynamics. Along this line of thought, it is rewarding that we can roughly get agreement with experiment (within errors) both for charmonium and botttomoniun using the same correcting factor $\sim 1/3.5$. This factor could be understood as evaluating the $\beta(\alpha_s)$ function at a low scale: $-\frac{2\pi\beta(\alpha_s)}{\beta_0\alpha_s^2} \sim 3.5$. Still, it is premature to draw any conclusion out of this. Note also that these estimates are significantly affected by the uncertainties of $\kappa$, and, for the $\pi^0$ case, by the uncertainty of $m_u/m_d$. On the other hand, the strong dependence on the light-quark masses makes these observables interesting for possible determination of the light-quark masses. 

For the ratios considered so far, Eqs.~(\ref{Rcchiral})-(\ref{Rbeta}), the prediction of the twist expansion is the same to the one we have found here. This is not so for the following ratio:
\begin{align}
\label{Rbcpipi}
R_{bc,\pi\pi}\equiv
R\left(\frac{\Upsilon(2S)\to \Upsilon(1S) \pi^+\pi^-}{\psi(2S)\to J/\psi \pi^+\pi^-}\right)
= & 6.65(^{+0.30}_{-0.38})_{\Lambda_1}(\mp 0.02)_{\kappa}(\pm 0.31)_{\rm s.p.}\times 10^{-2}
\,,
\\
\nn
R_{bc,\pi\pi}^{\rm exp}= & 5.59(0.50)\times 10^{-2}
\,.
\end{align}
In principle, we expect that for this ratio most of the neglected ${\cal O}(\alpha_s)$ corrections in the hadronization of the local operators vanish. This observable can be considered a rough measure of $\beta_{r,b}^{(21)}/\beta_{r,c}^{(21)}$. The agreement with experiment is remarkable: below 20\%, and could be accounted for by the quoted errors. 

We can also consider the following ratios:
 \begin{align}
 \label{Rbcpi}
R_{bc,\pi}\equiv &
R \left(\frac{\Upsilon(2S)\to \Upsilon(1S) \pi^0}{\psi(2S)\to J/\psi \pi^0} \right)=5.3(^{+0.2}_{-0.3})_{\Lambda_1}(\pm 0.25)_{\rm s.p.}\times 10^{-3} \,,\quad R_{bc,\pi}^{\rm exp}<3.4(1) \times 10^{-3}\,,\\
\label{Rbceta}
R_{bc,\eta}\equiv & R \left(\frac{\Upsilon(2S)\to \Upsilon(1S) \eta}{\psi(2S)\to J/\psi \eta} \right)=1.33(^{+0.06}_{-0.08})_{\Lambda_1 }(\pm 0.06)_{\rm s.p.}\times 10^{-3} \,,\quad R_{bc,\eta}^{\rm exp}=0.94(15) \times 10^{-3} \,.
\end{align}
Again, the twist expansion would yield a different prediction for these ratios, and, again, these ratios can be considered a measure of $\beta_{r,b}^{(21)}/\beta_{r,c}^{(21)}$. The agreement with experiment is quite reasonable, with difference of the order of 30\%. In order to assess whether this error and the error of \eq{Rbcpipi} are really related to $\beta_{r,b}^{(21)}/\beta_{r,c}^{(21)}$ or to something else, one may consider the double ratios
\begin{align}
\label{doubleratio}
&\frac{R_{bc,\pi}}{R_{bc,\pi\pi}}=\frac{R_{c,\pi}}{R_{b,\pi}}=0.08 \,, &&\frac{R^{\rm exp}_{c,\pi}}{R^{\rm exp}_{b,\pi}}<0.06(1) \,,\\
&\frac{R_{bc,\eta}}{R_{bc,\pi\pi}}=\frac{R_{c,\eta}}{R_{b,\eta}}= 2.0 \times 10 ^{-2}\,, &&\frac{R^{\rm exp}_{c,\eta}}{R^{\rm exp}_{b,\eta}}=1.7(3) \times 10 ^{-2}\,.
\end{align}
These double ratios are independent of $\beta_{r,b}^{(21)}/\beta_{r,c}^{(21)}$. We also expect several other uncertainties to cancel: the dependence on the light-quark masses cancels; in principle, we would also expect most of the effect due to unaccounted for ${\cal O}(\alpha_s)$ corrections in the hadronization of the local operators vanish, since the phase space is similar. An indirect reflection of this is that the dependence on $\kappa$ nearly vanishes. Nevertheless, this should be studied with more detail to make this statement more quantitative. For instance, for the $Q\bar Q(2S) \to Q\bar Q(1S)\eta$ transition, there is little phase space free. As a result, the phase space computation yields very different values for the bottomonium and charmonium transitions. Overall, we consider these double ratios as the cleanest objects to perform dedicated studies of the computational scheme discussed in this paper (and of the twist expansion, which yields the same prediction for these double ratios). For now, we just want to highlight that we find remarkably good agreement with experiment. This may indicate that the overlap of higher dimensional operators with the hybrids is small. In this respect, the possible incorporation of $B_c$ transitions to these analyses could be of much help.

\section{Conclusions}\label{conc}

We have studied one- and two-pion transitions between quarkonium states below threshold. We have used the weakly coupled version of pNRQCD, which assumes that $mv \gg \lQ$. On the other hand, our analysis does not need to constrain the relative size of $\lQ$ with respect to $mv^2$. For definiteness, we have generically considered the situation $mv^2 \ll \lQ$. This EFT of QCD for heavy quark-antiquark systems incorporates the $1/m_Q$ and multipole expansions systematically. We then write the hadronic representation for this EFT in terms of the singlet, representing the heavy quarkonium, hybrids and pion fields. In weakly coupled pNRQCD, the hybrid states consist of a color-octet heavy quark-antiquark field and a gluonic excitation operator, called gluelump, characterized by a $J^{PC}$. In order to explicitly compute the matching between both versions of the theory, we assume that the gluelump operators overlap predominantly with the lowest dimension gluonic operator with the same $J^{PC}$. The matching is completed by making use of low-energy theorems generated by the axial and energy-momentum tensor anomalies to obtain the local matrix elements for pion production by gluonic operators. 

In this framework, we have computed the $Q\bar{Q}(2S)\to Q\bar{Q}(1S)\pi\pi$ transitions and compared with the description obtained solely from chiral symmetry, as well as with experimental data. We do so for the normalized decay width spectra and for the total widths of charmonium and bottomonium. In the case of the normalized decay width spectra, both our approach and the purely chiral description fit the data well. However, the chiral description depends on two parameters that cannot be strongly constrained from the experimental data, whereas our approach only depends on one parameter $\kappa$. Our best fit to the combined charmonium and bottomonium data yields the value quoted in \eq{kappafinal} for $\kappa$.

Our computational scheme produces the same theoretical expression as the twist expansion \cite{Voloshin:1978hc} for the normalized decay width spectra computed in \Sec{Sec:Line-shape}. Note, however, that the twist expansion requires that $\lQ \ll mv^2$, and that the energy release $E \ll mv^2$, the latter condition is never fulfilled for two-pion transitions. Lower values for $\kappa$ than the ones in \eq{kappafinal} are found in \cite{Bai:1999mj,Chen:2019hmz}. The main source for the discrepancy is due to ${\cal O}(m_{\pi}^2)$ terms not included in these analysis. 

Our prediction for the total two-pion transition decay width depends on $\kappa$ [which we take from \eq{kappafinal}], and on $\beta_{r,Q}^{(12)}$. The latter coefficient suffers from large uncertainties. It involves a double sum: one over the gluelump states with the same quantum numbers and, for each gluelump, other sum over the states solution of its associated Schr\"odinger equation. In this paper, we have made a first estimate considering only the first gluelump and, for it, summing the first few states solution of the Schr\"odinger equation. Moreover, the states solution of the Schr\"odinger equation have a significant overlap with long distances. This makes the result quite dependent on the shape of the potentials. We have estimated this dependence considering different fit functions for the singlet and hybrid potentials. Overall, our estimates for the total decay widths $\Gamma_{\psi(2S)\to J/\psi\pi^+\pi^-}$, and $\Gamma_{\Upsilon(2S)\to\Upsilon(1S)\pi^+\pi^-}$ can be found in \eqs{Gammacpipi}{Gammabpipi}.

Enlarging our Lagrangian to include spin-dependent and isospin breaking operators we can use the same procedure to compute the width for $Q\bar{Q}(2S)\to Q\bar{Q}(1P)\pi^0$ transitions. Our prediction for the total width $\Gamma_{\psi(2S)\to h_c(1P)\pi^0}$ can be found in \eq{eq:hc}. This observable is interesting. Unlike previous decays, it does not suffer from the uncertainties associated to the hadronization of the local operator: the axial anomaly does not get ${\cal O}(\alpha)$ corrections, nor there are ${\cal O}(p^4)$ chiral corrections. We also observe a very weak dependence on variations of the wave function of the hybrid and singlet states: this object depends on $\beta_{\sigma}^{(12)}$. The major error is generated by the uncertainty on the energy difference between singlet and hybrid states. Once this error is taken into account the result is roughly compatible with experiment. 

Finally, we considered $Q\bar{Q}(2S)\to Q\bar{Q}(1S)\pi^0(\eta)$ transitions. Our results for the total decay widths $\Gamma_{\psi(2S)\to J/\psi \pi^0}$, $\Gamma_{\psi(2S)\to J/\psi \eta}$, $\Gamma_{\Upsilon(2S)\to \Upsilon(1S) \pi^0}$, and $\Gamma_{\Upsilon(2S)\to \Upsilon(1S) \eta}$ can be found in \eqss{Gamma2Spi1}{Gamma2Seta2}. 

Overall, we find that the $S$-wave hadronic transitions to P-wave heavy quarkonium are roughly compatible with theory within errors. For decays to $S$-wave heavy quarkonium,  the magnitude of our predictions is smaller than experiment, especially for the $Q\bar{Q}(2S)\to Q\bar{Q}(1S)\pi^0(\eta)$ decays.

We have also computed the ratios of the above transition rates in \Sec{Sec:Ratios}. We expect a more solid prediction for them. Depending on the ratio, different qualitative information on the theory can be obtained. The ratios of \eqs{Rcchiral}{Rbchiral} can be considered to be a test of chiral symmetry, as they are independent of the bound state dynamics, and there is no error associated to the hadronization of local gluonic operators. Good agreement with experiment is found. This may indicate a good behavior of chiral symmetry for these decays. Alternative ratios one may consider are \eqss{Rcpi}{Rbeta}. These ratios are also independent of the bound state dynamics but on the other hand suffer from unquantified errors due to the hadronization of the local gluonic operators. Large differences with experiment were found, which however could be accommodated by a single constant for bottomonium and charmonium. This agrees with expectations that the main corrections are independent of the bound state dynamics. These ratios and the ratios \eqs{Rcchiral}{Rbchiral} are independent of the bound state dynamics, i.e., they are independent of $\beta_{r,Q}^{(12)}$ and yield the same result as the twist expansion. On the other hand, the overall coefficients which, in principle, are calculable in both cases, are different for the total transition rates.

One may consider ratios that yield different predictions than the twist expansion. These are \eqss{Rbcpipi}{Rbceta}. We expect these ratios to be quite independent of the error associated to the hadronization of the local gluonic operator. On the other hand they depend on the bound state dynamics through the ratio: $\beta_{r,b}^{(12)}/\beta_{r,c}^{(12)}$. For \eq{Rbcpipi}, good agreement is found (below 20\%), and is quite reasonable for the other two ratios: of the order of 30\%. This may indicate that our evaluation of $\beta_{r,b}^{(12)}/\beta_{r,c}^{(12)}$ is quite reasonable.  Finally, we have considered double ratios in \eq{doubleratio}. In principle, these are the cleanest objects to compute [leaving aside \eqs{Rcchiral}{Rbchiral}]. They  are independent of the bound state dynamics, and we also expect them to be rather independent of the uncertainties associated to the hadronization of the local gluonic operator. The agreement with experiment is quite good for them.

We believe there is room for improvement, and leave for future work more dedicated analysis, in particular of the ratios and, specially, of the double ratio discussed in \eq{doubleratio}. This last object may allow a quantitative analysis of the validity of our computational scheme (and alternatively of the twist expansion). It would also be interesting to apply our computational scheme to the two- (or one) pion hadronic transition of the $B_c$. On a later stage, the possible incorporation of tetraquarks and B-meson loops for heavy quarkonium states near, or above threshold, should also be studied in greater detail. Actually, if enough precision is reached, discrepancies with experiment may hint at the need of incorporating those states.
\medskip

{\bf Acknowledgments} \\ 
We thank E. Guido and R. Mussa for sharing the experimental data of Ref.~\cite{Guido:2017cts} with us. We thank P. Petreczky for sharing the lattice data of Ref.~\cite{Bazavov:2017dsy} with us. We thank Feng-Kun Guo for providing us the data of \cite{Alexander:1998dq}. This work was supported in part by the Spanish Grants No. FPA2017-86989-P and No. SEV-2016-0588 from the Ministerio de Ciencia, Innovaci\'on y Universidades, and the Grant No. 2017SGR1069 from the Generalitat de Catalunya. J.T.C acknowledges the financial support from the European Union's Horizon 2020 research and innovation programme under the Marie Sk\l{}odowska--Curie Grant No. 665919.

\appendix

\section{Differential decay widths}\label{ddwap}

The most general amplitude for the two-pion transitions at ${\cal O}(p^2)$ in the chiral counting and using spin symmetry is given in \eq{2pgp1}. It is convenient to write it as 
\begin{align}
\mathcal{A}_{\pi\pi}&=(a_2-a_1)p^0_+p^0_--\frac{a_2}{2}m^2_{\pi\pi}+(a_2-a_3)m^2_{\pi}\,,\label{2pgp}
\end{align}
where $m_{\pi\pi}^2=(p_++p_-)^2$ is the dipion invariant mass, and $p_+$, $p_-$ refers to the momentum of the $\pi^+$ and $\pi^-$, respectively. 
In the reference frame of the decaying quarkonia, we find 
\begin{align}
p_+^0p_-^0=\frac{1}{4}\left(\Delta^2-\rho^2\sigma^2\cos^2\theta\right)\,,\label{epem}
\end{align}
with 
\begin{align}
\sigma&=\sqrt{1-4m_{\pi}^2/m_{\pi\pi}^2}\\
\Delta&=\frac{m^2_{H_i}-m^2_{H_f}+m^2_{\pi\pi}}{2m_{H_i}} \,,\label{rrep1}\\
\rho&=\frac{1}{2m_{H_i}}\sqrt{m^4_{H_i}+m^4_{H_f}+m^4_{\pi\pi}-2m^2_{H_i}m^2_{H_f}-2m^2_{H_i}m^2_{\pi\pi}-2m^2_{H_f}m^2_{\pi\pi}}\label{rrep2}\,,
\end{align}
 where $m_{H_i}$ and $m_{H_f}$ are the masses of initial and final quarkonium respectively. In the nonrelativistic approximation of the final quarkonium momentum, the above expressions reduce to
\begin{align}
&\Delta^{\rm nr}=\left(m_{H_i}-m_{H_f}\right)^2 \,,\label{rrep3}\\
&\rho^{\rm nr}=\sqrt{(m_{H_i}-m_{H_f})^2-m^2_{\pi\pi}}\label{rrep4}\,.
\end{align}
Using \eq{epem} in \eq{2pgp}, we can write
\begin{align}
\mathcal{A}_{\pi\pi}&=\frac{1}{4}(a_2-a_1)\left(\Delta^2-\rho^2\sigma^2\cos^2\theta\right)-\frac{a_2}{2}m^2_{\pi\pi}+(a_2-a_3)m^2_{\pi}\,.
\end{align}

The differential decay width is
\begin{align}
\frac{\Gamma_{\pi\pi}}{d m_{\pi\pi}d\cos\theta}&=\frac{\rho\sigma m_{\pi\pi}}{4(2\pi)^3}\left[\frac{1}{4}(a_2-a_1)\left(\Delta^2-\rho^2\sigma^2\cos^2\theta\right)-\frac{a_2}{2}m^2_{\pi\pi}+(a_2-a_3)m^2_{\pi}\right]^2\,,
\end{align}
and after integrating $\theta$ it reads
\begin{align}
\frac{d\Gamma_{\pi\pi}}{d m_{\pi\pi}}&=\frac{\rho\sigma m_{\pi\pi}}{4(2\pi)^3}\left[
2(a_2-a_3)^2m^4_{\pi}\left(1-\frac{2a_2}{a_2-a_3}\frac{m^2_{\pi\pi}}{4m^2_{\pi}}+\frac{a_2-a_1}{a_2-a_3}\frac{\Delta^2}{4m^2_{\pi}}\right)^2+\frac{(a_2-a_1)^2}{40}\rho^4\sigma^4 \right.\nn\\
&\left.-\frac{1}{3}(a_2-a_3)(a_2-a_1)m^2_{\pi}\rho^2\sigma^2\left(1-\frac{2a_2}{a_2-a_3}\frac{m^2_{\pi\pi}}{4m^2_{\pi}}+\frac{a_2-a_1}{a_2-a_3}\frac{\Delta^2}{4m^2_{\pi}}\right)\right]\,.\label{lnshp}
\end{align}

To obtain the total decay width we integrate numerically 
\begin{align}
\Gamma_{\pi\pi}=\int^{m_{H_i}-m_{H_f}}_{2m_{\pi}}d m_{\pi\pi}\frac{d\Gamma_{\pi\pi}}{d m_{\pi\pi}}\,.\label{lnshpint}
\end{align}

In the one-pion transition, the momenta are fixed by momentum conservation, in particular the final pion momentum is $|\bm{p}_\pi|=\rho(m_\pi)$, and the decay width is given by
\begin{align}
\Gamma_\pi=\frac{\rho(m_{\pi})}{2\pi}|\mathcal{A_\pi}|^2\,,\label{dw1p}
\end{align}
with $\rho(m_\pi)$ is given in \eq{rrep2} replacing $m_{\pi\pi}$ by $m_{\pi}$.

\section{Singlet and hybrid bound states}
\label{AppStates}

The heavy quarkonium wave function reads
\begin{align}
S^{n j m_j l s }(\bm{r})=\phi^{(n)}(r)\sum_{m_l m_s} \mathcal{C}^{j m_j}_{l\,m_l\,s\,m_s}Y_{l\,m_l}(\theta,\phi)\chi_{s\,m_s}\equiv\phi^{(n)}(r)\Phi^0_{^{2s+1}l_j}(\theta,\phi)\,,
\end{align}
where $\phi^{(n)}(r)$ is the solution of 
\begin{align}
 \left[-\frac{1}{m_Qr^2}\,\partial_r\,r^2\,\partial_r+\frac{l(l+1)}{m_Qr^2}+ V_{\Sigma_g^+}(r)\right]\phi^{(n)}(r)=
 \mathcal{E}_n\,\phi^{(n)}(r)\,.\label{SchSinglet}
\end{align}

\begin{figure}[ht!]
\includegraphics[width=.6\textwidth]{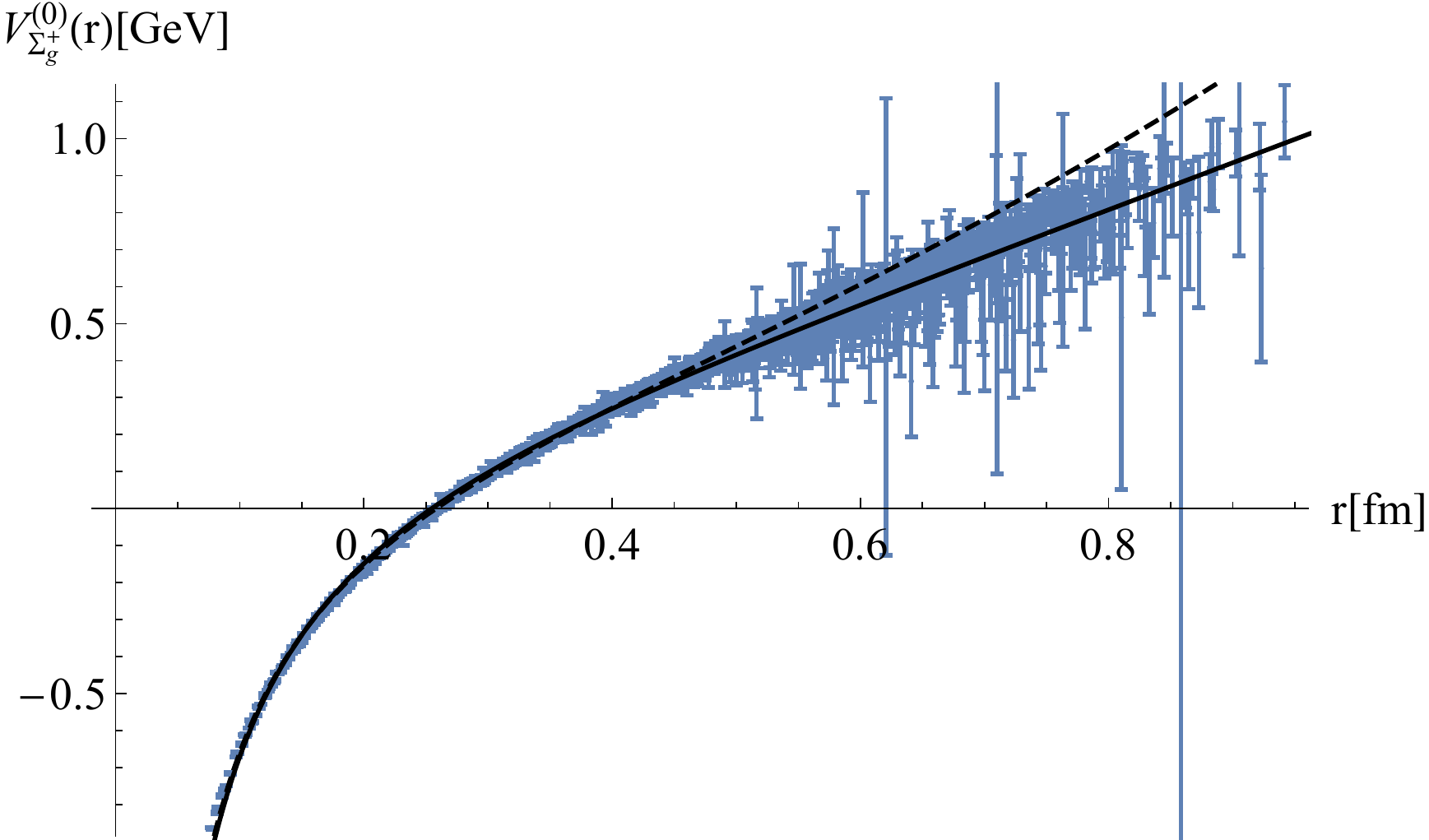}\\
	\caption{The static singlet potential in Eqs.~\eqref{VSingletfl} and \eqref{VSinglet} fitted to the lattice data of Ref.~\cite{Bazavov:2017dsy} (blue dots) as solid and dashed lines, respectively.}
	\label{p0pm}
 \end{figure}

It is not our aim in this paper to make a detailed analysis and optimization of the solutions of the bound states but rather to make a qualitative analysis. For the numerical evaluations of the hybrid and quarkonium spectra, we use the values $m_c(1{\rm GeV})=1.496$ GeV and $m_b(1{\rm GeV})=4.885$ GeV \cite{Peset:2018ria} for the heavy quark masses in the RS$^{\prime}$ scheme \cite{Pineda:2001zq}. To explore the sensitivity of the matrix elements to the shape of the potential, we have considered two versions of the potentials. The first is the potential in \eq{VSinglet} with the nonperturbative constant $b_{\Sigma_g^+}$ obtained by fitting to the lattice data from Ref.~\cite{Bazavov:2017dsy} up to $r=1$ fm together with a free energy offset, $b^{\rm offset}_{\Sigma_g^+}$. The second version of the potential is a fit of the same lattice data to a function constrained to reproduce \eq{VSinglet} in the short distance and a to have a linear behavior in the long distance, while overall adjusting well to the data. The function chosen is
\begin{align}
V_{\Sigma_g^+}=V_s^{(0)}(r)+\left(c^{(0)}_{\Sigma_g^+}+c^{(1)}_{\Sigma_g^+}r^2+c^{(2)}_{\Sigma_g^+}r^4\right)^{1/4}\,.\label{VSingletfl}
\end{align}
For the perturbative static singlet potential we take the $\mathcal{O}(\alpha^3_s)$ result\footnote{The ${\cal O}(\alpha_s)$ term was computed in \cite{Fischler:1977yf}, the ${\cal O}(\alpha_s^2)$  in \cite{Schroder:1998vy,Peter:1996ig}, the ${\cal O}(\alpha^3_s)$ logarithmic term in \cite{Brambilla:1999qa}, the light-flavor finite piece in \cite{Smirnov:2008pn}, and the pure gluonic finite piece in \cite{Anzai:2009tm,Smirnov:2009fh}.} evaluated at $\nu = 5.6$ GeV (the shortest available scale). The parameters obtained from the fits are the following
\begin{align}
&b_{\Sigma_g^+}=4.07\cdot10^{-2}~{\rm GeV}^3\,,\quad b^{\rm offset}_{\Sigma_g^+}=0.695~{\rm GeV}\,,\\
&c^{(0)}_{\Sigma_g^+}=0.222~{\rm GeV}^4\,,\quad c^{(1)}_{\Sigma_g^+}=6.91\cdot 10^{-2}~{\rm GeV}^6\,,\quad c^{(2)}_{\Sigma_g^+}=3.54\cdot 10^{-3}~{\rm GeV}^8\,.
\end{align}
In Fig.~\ref{p0pm}, we plot the lattice data together with the fitted \eq{VSinglet} (with the offset) and \eq{VSingletfl} potentials as dashed and solid lines, respectively. The spectra obtained are displayed in Tables~\ref{qecp1} and~\ref{qecp2} for the potential in Eqs.~\eqref{VSinglet} and  in \eqref{VSingletfl}, respectively. The origin of energies is chosen in order to reproduce the experimental spin average of the ground state.

\begin{table}[ht!]
 \begin{tabular}{||c|c||c|c|c||c|c|c||}
 \hline\hline
 \multirow{2}{*}{nl} & \multirow{2}{*}{$J^{PC}$} & \multicolumn{3}{c||}{$c\bar{c}$} & \multicolumn{3}{c||}{$b\bar{b}$}\\ \cline{3-8}
                     &                  & $m_H$ & $\langle1/r\rangle$ & $E_{kin}$ & $m_H$ & $\langle1/r\rangle$ & $E_{kin}$ \\ \hline
 
 1S & $\{0^{-+},1^{--}\}$       & 3.068 & 0.66 & 0.31 & 9.442  & 1.15 & 0.28 \\ \hline
 2S & $\{0^{-+},1^{--}\}$       & 3.686 & 0.43 & 0.48 & 9.916  & 0.67 & 0.36 \\ \hline
 1P & $\{1^{+-},(0,1,2)^{++}\}$ & 3.464 & 0.37 & 0.39 & 9.779  & 0.59 & 0.30 \\ \hline
 1D & $\{2^{-+},(1,2,3)^{--}\}$ & 3.775 & 0.27 & 0.48 & 10.009 & 0.42 & 0.35 \\ \hline\hline
 \end{tabular}
 \caption{Quarkonium spectrum obtained from the  potential in \eq{VSingletfl} fitted to the lattice data Ref.~\cite{Bazavov:2017dsy}. The origin of energies is chosen in order to reproduce the spin average of the ground state. All dimension-full entries are in GeV.}
\label{qecp1}
\end{table}
\begin{table}[ht!]
 \begin{tabular}{||c|c||c|c|c||c|c|c||}
 \hline\hline
 \multirow{2}{*}{nl} & \multirow{2}{*}{$J^{PC}$} & \multicolumn{3}{c||}{$c\bar{c}$} & \multicolumn{3}{c||}{$b\bar{b}$}\\ \cline{3-8}
                     &                  & $m_H$ & $\langle1/r\rangle$ & $E_{kin}$ & $m_H$ & $\langle1/r\rangle$ & $E_{kin}$ \\ \hline
 
 1S & $\{0^{-+},1^{--}\}$       & 3.068 & 0.73 & 0.38 & 9.442  & 1.36 & 0.38 \\ \hline
 2S & $\{0^{-+},1^{--}\}$       & 3.847 & 0.51 & 0.67 & 10.015 & 0.76 & 0.45 \\ \hline
 1P & $\{1^{+-},(0,1,2)^{++}\}$ & 3.553 & 0.42 & 0.50 & 9.873  & 0.64 & 0.35 \\ \hline
 1D & $\{2^{-+},(1,2,3)^{--}\}$ & 3.949 & 0.32 & 0.65 & 10.145 & 0.46 & 0.41\\ \hline\hline
 \end{tabular}
 \caption{Quarkonium spectrum obtained from the potential in \eq{VSinglet} fitted to the lattice data Ref.~\cite{Bazavov:2017dsy}. The origin of energies is chosen in order to reproduce the spin average of the ground state. All dimension-full entries are in GeV.}
\label{qecp2}
\end{table}

The hybrid wave functions are obtained following the procedure described in Ref.~\cite{Berwein:2015vca}, by solving the coupled Schr\"odinger equations involving the potentials $V^{(0)}_{\Sigma^-_u}(r)-V^{(0)}_{\Pi_u}(r)$ or $V^{(0)}_{\Sigma^{+\prime}_g}(r)-V^{(0)}_{\Pi_g}(r)$ generated by the $1^{+-}$ and $1^{--}$ gluelump at short distances, respectively [see \eq{Vkappa}]. There are two types of solution of $\Psi_{1\lambda}$ corresponding to states with opposite parity $\left(\Psi^{m j m_j \ell s }_{+}\right)_{\la}$ and $\left(\Psi^{m j m_j \ell s }_{-}\right)_{\la}$,

\begin{align}
&\Psi^{m j m_j \ell s }_{+}(\bm{r})=\sum_{m_\ell m_s} \mathcal{C}^{j m_j}_{l\,m_\ell\,s\,m_s}\left(
\begin{array}{c}
\psi_0^{(m)}(r)v_{\ell\,m_\ell}^0(\theta,\phi) \\
 \frac{1}{\sqrt{2}}\psi_{+}^{(m)}(r)v_{\ell\,m_\ell}^{+1}(\theta,\phi) \\
\frac{1}{\sqrt{2}}\psi_{+}^{(m)}(r)v^{-1}_{\ell\,m_\ell}(\theta,\phi) \\
\end{array}\right)\chi_{s\,m_s}=\left(\begin{array}{c}
\psi_0^{(m)}(r)\Phi^0_{^{2s+1}l_j}(\theta,\phi) \\
 \frac{1}{\sqrt{2}}\psi_{+}^{(m)}(r)\Phi^{+1}_{^{2s+1}\ell_j}(\theta,\phi) \\
\frac{1}{\sqrt{2}}\psi_{+}^{(m)}(r)\Phi^{-1}_{^{2s+1}\ell_j}(\theta,\phi) \\
\end{array}\right)
\,,\label{psip} \\
&\Psi^{m j m_j \ell s }_{-}(\bm{r})=\sum_{m_\ell m_s} \mathcal{C}^{j m_j}_{\ell\,m_\ell\,s\,m_s}\left(
\begin{array}{c}
 0 \\
 \frac{1}{\sqrt{2}}\psi_{-}^{(m)}(r)v_{\ell\,m_\ell}^{+1}(\theta,\phi) \\
 -\frac{1}{\sqrt{2}}\psi_{-}^{(m)}(r)v_{\ell\,m_\ell}^{-1}(\theta,\phi) \\
\end{array}
\right)\chi_{s\,m_s}=\left(
\begin{array}{c}
 0 \\
 \frac{1}{\sqrt{2}}\psi_{-}^{(m)}(r)\Phi^{+1}_{^{2s+1}\ell_j}(\theta,\phi) \\
 -\frac{1}{\sqrt{2}}\psi_{-}^{(m)}(r)\Phi^{-1}_{^{2s+1}\ell_j}(\theta,\phi) \\
\end{array}
\right)\,.
\label{psim}
\end{align}

The parity and charge conjugation of these states corresponding to the $\Psi_{\pm}$ solutions are
\begin{align}
P=\pm(-1)^\ell P_G\,,\quad C=\mp(-1)^{\ell+s}C_G\,,
\end{align}
with $P_G$ and $C_G$ the parity and charge conjugation of the gluelump associated to the states. 

The angular wave functions of the hybrid states are the eigenfunctions of
\begin{align}
\left(\bm{L}^2_{\bar{Q}Q}+\frac{\la^2}{\sin^2\theta}+2\la\frac{\cos\theta}{\sin^2\theta}i\pa_\theta\right)v_{\ell\,m_\ell}^{\la}=\ell(\ell+1)v_{\ell\,m_\ell}^{\la}\,,
\end{align}
which for $\la=0$ corresponds to the usual differential equation for spherical harmonics and, therefore, $v_{l\,m_l}^{0}=Y_{l\,m_l}$. Note that we have chosen to represent the angular momentum quantum numbers of the hybrids and quarkonia by $\ell$ and $l$, respectively to highlight the difference in angular wave functions. When using spectroscopic notation to specify the quarkonium states we use the usual $S,P,D,\dots$ for $l=0,1,2,\dots,$ and for hybrid states, we use $\mathcal{S},\mathcal{P},\mathcal{D},\dots$ for $\ell=0,1,2,\cdots$.

The $\la^{2s+1}\ell_j$ spin-angular hybrid wave functions can be shown to be
\begin{align}
\Phi^{\la}_{^1\mathcal{S}_0}(\theta,\phi)&=\frac{1}{\sqrt{4\pi}}\frac{\mathbbm{1}}{\sqrt{2}}\de_{\la 0}\,,\label{saw1}\\
\Phi^{\la}_{^3\mathcal{S}_1}(\theta,\phi)&=\frac{1}{\sqrt{4\pi}}\frac{\bm{\sigma}\cdot\hat{e}_{m_J}}{\sqrt{2}}\de_{\la 0}\,,\label{saw2}\\
\Phi^{\la}_{^1\mathcal{P}_1}(\theta,\phi)&=\sqrt{\frac{3}{4\pi}}\hat{\bm{r}}^{i\dag}_{\la}\hat{e}^i_{m_J} \frac{\mathbbm{1}}{\sqrt{2}}\,,\label{saw3}\\
\Phi^{\la}_{^3\mathcal{P}_0}(\theta,\phi)&=\frac{1}{\sqrt{8\pi}}\hat{\bm{r}}^{i\dag}_{\la}\bm{\sigma}^i\,,\label{saw4}\\
\Phi^{\la}_{^3\mathcal{P}_1}(\theta,\phi)&=i\sqrt{\frac{3}{8\pi}}\frac{\bm{\sigma}\cdot(\hat{\bm{r}}^{\dag}_{\la}\times\hat{e}_{m_J})}{\sqrt{2}}\,,\label{saw5}\\
\Phi^{\la}_{^3\mathcal{P}_2}(\theta,\phi)&=\sqrt{\frac{3}{8\pi}}\bm{\sigma}^ih^{ij}_{2m_J}\hat{\bm{r}}^{j\dag}_{\la}\,,\label{saw6}
\end{align}
where $\hat{e}$ are the polarization vectors
\begin{align}
&\hat{e}_0=(0,\,0,\,1)\,,\\
&\hat{e}_{\pm}=\frac{\mp1}{\sqrt{2}}(1,\,\pm i,\,0)\,,
\end{align}
and the tensors $h^{ij}_{2m_J}$ are traceless, completely symmetric, and normalized as
\begin{align}
h^{*ij}_{2m_J}h^{ij}_{2m_J^{\prime}}=\delta_{m_Jm_J^{\prime}}\,.
\end{align}

The $^{2s+1}l_j$ spin-angular wave functions for quarkonium correspond to the $\la=0$, $^{2s+1}\ell_j$ hybrids wave functions.

The radial wave function in \eq{psip} is obtained by (numerically) solving the following coupled radial Schr\"odinger equations:
\begin{align}
 \hspace{-4pt}\left[-\frac{1}{m_Q r^2}\,\partial_rr^2\partial_r+\frac{1}{m_Qr^2}\begin{pmatrix} \ell(\ell+1)+2 & -2\sqrt{\ell(\ell+1)} \\ -2\sqrt{\ell(\ell+1)} & \ell(\ell+1) \end{pmatrix}+\begin{pmatrix} V_{\Sigma} & 0 \\ 0 & V_{\Pi} \end{pmatrix}\right]\hspace{-4pt}\begin{pmatrix} \psi_0^{(m)} \\ \psi_{+}^{(m)}\end{pmatrix}=\mathcal{E}_m\begin{pmatrix} \psi_0^{(m)} \\ \psi_{+}^{(m)}\end{pmatrix}\,.
\end{align}
For the special case $\ell=0$ the equations decouple and in fact $\psi_{+}^{(m)}$ becomes irrelevant since $v^{\pm 1}_{00}$ does not exist. The radial wave function in \eq{psim} is obtained from the uncoupled radial Schr\"odinger equation
\begin{align}
 \left[-\frac{1}{m_Qr^2}\,\partial_r\,r^2\,\partial_r+\frac{\ell(\ell+1)}{m_Qr^2}+ V_{\Pi}\right]\psi_{-}^{(m)}=\mathcal{E}_m\,\psi_{-}^{(m)}\,.\label{ucse}
\end{align}

The hybrid static potentials, $V_{\Pi_u}$, $V_{\Sigma^-_u}$, $V_{\Pi_g}$, and  $V_{\Sigma^{+\prime}_g}$, match to the NRQCD heavy quark-antiquark static energies, which have been computed on the lattice \cite{Juge:2002br,Bali:2000vr,Capitani:2018rox} and are given up to NLO in the multipole expansion by the potential in \eq{Vkappa}. As in the static potential for the singlet, we have considered two versions of the potentials in order to assess the sensitivity of the matrix elements to the shape of the potential. The first is the potential in \eq{Vkappa} with the nonperturbative constant $b_{k\la}$ obtained by fitting to the lattice data from Ref.~\cite{Juge:2002br} up to $r=1$~fm together with a free energy offset. The second version of the potential is a fit of the same lattice data up to $r\leq 2$~fm by a function constrained to reproduce \eq{Vkappa} in the short distance and a to have a linear behavior in the long distance while overall adjusting well to the lattice data. The function is
\begin{align}
V_{\Lambda^{\sigma}_\eta}=V_o^{(0)}(r)+\left(c^{(0)}_{\Lambda^{\sigma}_\eta}+c^{(1)}_{\Lambda^{\sigma}_\eta}r^2+c^{(2)}_{\Lambda^{\sigma}_\eta}r^4\right)^{1/4}\,.\label{Vkappafl}
\end{align}
The perturbative static octet potential is taken at $\mathcal{O}(\alpha^3_s)$ from Ref.~\cite{Kniehl:2004rk}, and evaluated at $\nu = 2.6$ GeV (the shortest available scale). The results for the hybrid spectrum for $V_{\Pi_u}$-$V_{\Sigma^-_u}$  associated to the $1^{+-}$ gluelump are discussed at length in Ref.~\cite{Berwein:2015vca}.\footnote{In Ref.~\cite{Berwein:2015vca}, the choice of \eq{Vkappafl} was different but the resulting potentials are completely equivalent in practice.} The parameters obtained from the fits are the following:
\begin{align}
&b_{\Sigma^{+\prime}_g}=1.26\cdot 10^{-2}~{\rm GeV}^3\,,\quad b^{\rm offset}_{\Sigma^{+\prime}_g}=1.22~{\rm GeV}\,,\\
&b_{\Pi_g}=1.93\cdot 10^{-2}~{\rm GeV}^3\,,\quad b^{\rm offset}_{\Pi_g}=1.22~{\rm GeV}\,,\\
&c^{(0)}_{\Sigma^{+\prime}_g}=1.58~{\rm GeV}^4\,,\quad c^{(1)}_{\Sigma^{+\prime}_g}=12.9\cdot 10^{-2}~{\rm GeV}^6\,,\quad c^{(2)}_{\Sigma^{+\prime}_g}=1.94\cdot 10^{-3}~{\rm GeV}^8\,,\\
&c^{(0)}_{\Pi_g}=1.77~{\rm GeV}^4\,,\quad c^{(1)}_{\Pi_g}=10.6\cdot 10^{-2}~{\rm GeV}^6\,,\quad c^{(2)}_{\Pi_g}\equiv c^{(2)}_{\Sigma^{+\prime}_g}\,.
\end{align}

\begin{figure}[ht!]
\includegraphics[width=.6\textwidth]{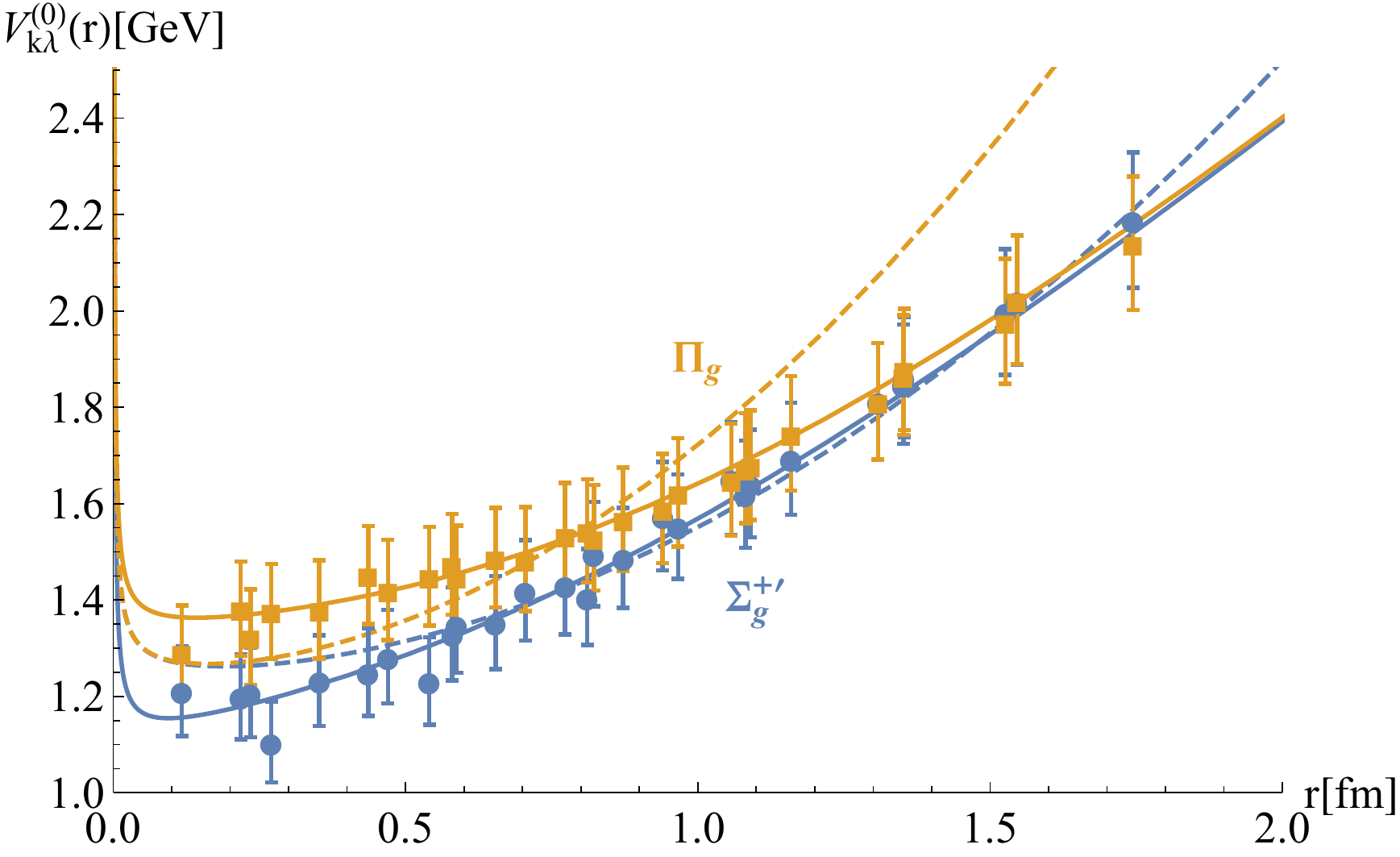}\\
\caption{Lattice data for the static energies $\Sigma^{+\prime}_g$ and $\Pi_g$ from Ref.~\cite{Juge:2002br} associated to the gluelump $k^{PC}=1^{--}$ plotted as blue dots and orange squares, respectively. The blue (orange) solid and dashed lines are the $\Sigma^{+\prime}_g$($\Pi_g$) potentials of Eqs.~\eqref{VSingletfl} and \eqref{VSinglet} fitted to the lattice data, respectively.}
\label{p1mm}
\end{figure}

The fits of the potentials $V_{\Pi_g}$-$V_{\Sigma^{+\prime}_g}$, associated to the $1^{--}$ gluelump, are shown in Fig.~\ref{p1mm} in solid and dashed lines and the corresponding spectra in Tables~\ref{meth11} and ~\ref{meth12} for the potentials in Eqs.~\eqref{Vkappafl} and \eqref{Vkappa}, respectively. The origin of energies is chosen, as in Ref.~\cite{Berwein:2015vca}, such that the short distance limit of $V_{\Pi_u}$ and $V_{\Sigma^-_u}$ tend to $V_o^{0}+\Lambda_{1^{+-}}$ with the value of $\Lambda_{1^{+-}}=0.87(15)$~GeV computed in Ref.~\cite{Bali:2003jq}.

\begin{table}[ht!]
 \begin{tabular}{||c|c|c|c||c|c|c|c||c|c|c|c||}
 \hline\hline
 \multirow{2}{*}{\text{multiplet}} &  \multirow{2}{*}{l} & \multirow{2}{*}{$J^{PC}$}  & \multirow{2}{*}{$\Lambda^{\sigma}_{\eta}$} & \multicolumn{4}{c||}{$c\bar{c}$} & \multicolumn{4}{c||}{$b\bar{b}$}\\
 \cline{5-8}\cline{9-12}
 & & & & $m_H$ & $\langle1/r\rangle$ & $E_{kin}$ & $P_\Pi$ & $m_H$ & $\langle1/r\rangle$ & $E_{kin}$ & $P_\Pi$ \\
\hline
 $H_1(1\mathcal{P})$  & \multirow{4}{*}{1} & \multirow{2}{*}{$\{1^{+-},(0,1,2)^{++}\}$} & \multirow{2}{*}{$\Sigma_g^{+\prime}$-$\Pi_g$} & 4.51 & 0.42 & 0.15 & 0.57 & 11.14 & 0.55 & 0.09 & 0.44 \\ \cline{5-12}
 $H_1'(2\mathcal{P})$ &                    &                                            &                                               & 4.86 & 0.27 & 0.33 & 0.20 & 11.33 & 0.40 & 0.19 & 0.27 \\ \cline{1-1}\cline{3-12}
 $H_2(1\mathcal{P})$  &                    & \multirow{2}{*}{$\{1^{-+},(0,1,2)^{--}\}$} & \multirow{2}{*}{$\Pi_g$}                      & 4.74 & 0.27 & 0.22 & 1.00 & 11.30 & 0.38 & 0.13 & 1.00 \\ \cline{5-12}
 $H_2'(2\mathcal{P})$ &                    &                                            &                                               & 5.01 & 0.24 & 0.37 & 1.00 & 11.51 & 0.34 & 0.26 & 1.00 \\ \hline
 $H_3(1\mathcal{S})$  & \multirow{2}{*}{0} & \multirow{2}{*}{$\{0^{-+},1^{--}\}$}       & \multirow{2}{*}{$\Sigma_g^{+\prime}$}         & 4.63 & 0.29 & 0.25 & 0.00 & 11.19 & 0.40 & 0.15 & 0.00 \\ \cline{5-12}
 $H_3'(2\mathcal{S})$ &                    &                                            &                                               & 5.01 & 0.25 & 0.40 & 0.00 & 11.41 & 0.35 & 0.24 & 0.00 \\ \hline
 $H_4(1\mathcal{D})$  & \multirow{4}{*}{2} & \multirow{2}{*}{$\{2^{-+},(1,2,3)^{--}\}$} & \multirow{2}{*}{$\Sigma_g^{+\prime}$-$\Pi_g$} & 4.70 & 0.28 & 0.24 & 0.50 & 11.24 & 0.38 & 0.14 & 0.39 \\ \cline{5-12}
 $H_4'(2\mathcal{D})$ &                    &                                            &                                               & 5.06 & 0.23 & 0.42 & 0.24 & 11.44 & 0.32 & 0.25 & 0.26 \\ \cline{1-1}\cline{3-12}
 $H_5(1\mathcal{D})$  &                    & \multirow{2}{*}{$\{2^{+-},(1,2,3)^{++}\}$} & \multirow{2}{*}{$\Pi_g$}                      & 4.92 & 0.22 & 0.32 & 1.00 & 11.40 & 0.30 & 0.18 & 1.00 \\ \cline{5-12}
 $H_5'(2\mathcal{D})$ &                    &                                            &                                               & 5.27 & 0.19 & 0.47 & 1.00 & 11.61 & 0.28 & 0.28 & 1.00 \\ \hline\hline
 \end{tabular}
 \caption{Heavy hybrid spectrum for $k^{PC}=1^{--}$ obtained from the fit of \eq{VSingletfl} to the lattice data of Ref.~\cite{Juge:2002br}. All dimension-full entries are in GeV. $P_{\Pi}=\int d^3r |\psi_{+/-}^{(m)}|^2$ is the integral over the square of the wave function associated with the $\Pi_g$ potential. It can be interpreted as the probability to find the hybrid in a $\Pi_g$ configuration, thus it gives a measure of the mixing effects.}
\label{meth11}
\end{table}

\begin{table}[ht!]
 \begin{tabular}{||c|c|c|c||c|c|c|c||c|c|c|c||}
 \hline\hline
 \multirow{2}{*}{\text{multiplet}} &  \multirow{2}{*}{l} & \multirow{2}{*}{$J^{PC}$}  & \multirow{2}{*}{$\Lambda^{\sigma}_{\eta}$} & \multicolumn{4}{c||}{$c\bar{c}$} & \multicolumn{4}{c||}{$b\bar{b}$}\\
 \cline{5-8}\cline{9-12}
 & & & & $m_H$ & $\langle1/r\rangle$ & $E_{kin}$ & $P_\Pi$ & $m_H$ & $\langle1/r\rangle$ & $E_{kin}$ & $P_\Pi$ \\
\hline
 
 $H_1(1\mathcal{P})$  & \multirow{4}{*}{1} & \multirow{2}{*}{$\{1^{+-},(0,1,2)^{++}\}$} & \multirow{2}{*}{$\Sigma_g^{+\prime}$-$\Pi_g$} & 4.54 & 0.44 & 0.15 & 0.59 & 11.18 & 0.57 & 0.08 & 0.58 \\ \cline{5-12}
 $H_1'(2\mathcal{P})$ &                    &                                            &                                               & 4.87 & 0.28 & 0.32 & 0.20 & 11.36 & 0.37 & 0.18 & 0.19 \\ \cline{1-1}\cline{3-12}
 $H_2(1\mathcal{P})$  &                    & \multirow{2}{*}{$\{1^{-+},(0,1,2)^{--}\}$} & \multirow{2}{*}{$\Pi_g$}                      & 4.79 & 0.31 & 0.28 & 1.00 & 11.32 & 0.41 & 0.15 & 1.00 \\ \cline{5-12}
 $H_2'(2\mathcal{P})$ &                    &                                            &                                               & 5.24 & 0.28 & 0.51 & 1.00 & 11.57 & 0.37 & 0.28 & 1.00 \\ \hline
 $H_3(1\mathcal{S})$  & \multirow{2}{*}{0} & \multirow{2}{*}{$\{0^{-+},1^{--}\}$}       & \multirow{2}{*}{$\Sigma_g^{+\prime}$}         & 4.68 & 0.28 & 0.23 & 0.00 & 11.26 & 0.37 & 0.12 & 0.00 \\ \cline{5-12}
 $H_3'(2\mathcal{S})$ &                    &                                            &                                               & 5.05 & 0.25 & 0.41 & 0.00 & 11.46 & 0.33 & 0.23 & 0.00 \\ \hline
 $H_4(1\mathcal{D})$  & \multirow{4}{*}{2} & \multirow{2}{*}{$\{2^{-+},(1,2,3)^{--}\}$} & \multirow{2}{*}{$\Sigma_g^{+\prime}$-$\Pi_g$} & 4.74 & 0.29 & 0.26 & 0.46 & 11.29 & 0.39 & 0.14 & 0.46 \\ \cline{5-12}
 $H_4'(2\mathcal{D})$ &                    &                                            &                                               & 5.06 & 0.23 & 0.42 & 0.25 & 11.47 & 0.30 & 0.23 & 0.25 \\ \cline{1-1}\cline{3-12}
 $H_5(1\mathcal{D})$  &                    & \multirow{2}{*}{$\{2^{+-},(1,2,3)^{++}\}$} & \multirow{2}{*}{$\Pi_g$}                      & 5.02 & 0.25 & 0.39 & 1.00 & 11.44 & 0.33 & 0.22 & 1.00 \\ \cline{5-12}
 $H_5'(2\mathcal{D})$ &                    &                                            &                                               & 5.47 & 0.23 & 0.62 & 1.00 & 11.69 & 0.31 & 0.34 & 1.00 \\ \hline\hline
 \end{tabular}
 \caption{Heavy hybrid spectrum for $k^{PC}=1^{--}$ obtained from the fit of \eq{VSinglet} to the lattice data of Ref.~\cite{Juge:2002br}. All dimension-full entries are in GeV. $P_{\Pi}=\int d^3r |\psi_{+/-}^{(m)}|^2$ is the integral over the square of the wave function associated with the $\Pi_g$ potential. It can be interpreted as the probability to find the hybrid in a $\Pi_g$ configuration, thus it gives a measure of the mixing effects.}
\label{meth12}
\end{table}

\section{Matrix Elements}\label{app:matelem}

We now compute the matrix elements that appear in the hadronic transitions. We suppress the labels $k^{PC}=1^{+-}$ and $k^{PC}=1^{--}$ on the hybrid fields since it does not affect the calculation. For simplicity, we denote the quarkonia and hybrid wave functions just as $S_n$ and $\Psi_m$, respectively:
\begin{align}
\langle\Psi_m|
\hat{\bm{r}}^{\dag}_{\la}\cdot\bm{r}
|S_n\rangle&=\sum_{m_\ell m^{\prime}_s m_l m_s} \mathcal{C}^{j m_j}_{\ell\,m_\ell\,s^{\prime}\,m^{\prime}_s}\mathcal{C}^{j m_j}_{l\,m_l\,s\,m_s}\de_{ss^{\prime}}\de_{m_sm_s^{\prime}}\int dr\,r^3 \psi_{0}^{(m)\dag}(r)\phi^{(m)}(r)\nn\\
&\times\int d\Omega\,  v_{\ell\,m_\ell}^{0\dag}(\theta,\phi)Y_{l\,m_l}(\theta,\phi)\,,\label{me1a}
\end{align}
where we have used that $\hat{\bm{r}}^{\dag}_{\la}\cdot\bm{r}=r\de_{\la 0}$. For $\la=0$ $v^{0}_{lm}=Y_{l\,m_l}$ so we can use the spherical harmonics orthogonality relations
\begin{align}
\eqref{me1a}&=\de_{ss^{\prime}}\sum_{m^{\prime}_lm_l m_s} \mathcal{C}^{j m_j}_{\ell\,m_\ell\,s\,m_s}\mathcal{C}^{j m_j}_{l\,m_l\,s\,m_s}\de_{l\ell}\de_{m_lm_\ell}\int dr\,r^3 \psi_{0}^{(m)\dag}(r)\phi^{(n)}(r) \nn\\
&=\de_{ss^{\prime}}\de_{l\ell}\sum_{m_s} \left(\mathcal{C}^{j m_j}_{l\,m_j-m_s\,s\,m_s}\right)^2\int dr\,r^3 \psi_{0}^{(m)\dag}(r)\phi^{(n)}(r)=\de_{ss^{\prime}}\de_{l\ell}\int dr\,r^3 \psi_{0}^{(m)\dag}(r)\phi^{(n)}(r)\,.\label{me1}
\end{align}

Let us work out some specific matrix elements necessary for the $\psi(2S)\to h_c\pi^0$ transition,
\begin{align}
\langle\Psi(m^3\mathcal{S}_1)| 
\hat{\bm{r}}^{\dag}_{\la}\cdot( \bm{\sigma}_1-\bm{\sigma}_2)
|S(n^1P_1)\rangle&=\int d\Omega\frac{\sqrt{3}}{2\pi}\hat{e}_{m^{\prime}_J}^{j*}\hat{\bm{r}}^{j}_{0} \hat{\bm{r}}^{i}_{0}\hat{e}^i_{m_J} \int dr\, r^2 \psi_{0}^{{(m)}\dag}(r)\phi^{(n)}(r)\nn\\
&=\frac{2}{\sqrt{3}}\delta_{m_J m^{\prime}_J}\int dr\, r^2 \psi_{0}^{{(m)}\dag}(r)\phi^{(n)}(r)\,,
\end{align}
\begin{align}
\langle\Psi(m^1\mathcal{P}_1)| 
\hat{\bm{r}}^{\dag}_{\la}\cdot( \bm{\sigma}_1-\bm{\sigma}_2)
|S(n^3S_1)\rangle&=\sum_{\la}\int d\Omega\frac{\sqrt{3}}{2\pi}\hat{e}_{m^{\prime}_J}^{j*}\hat{\bm{r}}^{j}_{\la} \hat{\bm{r}}^{i}_{\la}\hat{e}^i_{m_J} \int dr\, r^2 \psi_{\la}^{{(m)}\dag}(r)\phi^{(n)}(r)\nn\\
&=\sum_{\la}\frac{2}{\sqrt{3}}\delta_{m_J m^{\prime}_J}\int dr\, r^2 \psi_{\la}^{{(m)}\dag}(r)\phi^{(n)}(r)\,,
\end{align}
where we have used that hybrid states in $\ell=0$ must have $\la=0$ and also that
\begin{align}
\int d\Omega \hat{\bm{r}}^{i\,\dag}_{\la}\hat{\bm{r}}^{j}_{\la}=\frac{4\pi}{3}\delta^{ij}\,.\label{agint}
\end{align}
For $P$-wave pion (or eta) emission, we need the following matrix element:
\begin{align}
&\langle\Psi(m^3\mathcal{S}_1)|\left(\bm{\sigma}_1+\bm{\sigma}_2\right)\cdot\hat{\bm{r}}_{\la}^{\dag}\bm{r}^j|S(n^3S_1)\rangle=\frac{1}{8\pi}\hat{e}_{m^{\prime}_J}^{l*}\hat{e}_{m_J}^{k}{\rm Tr}\left[\left[\bm{\sigma}^i,\,\bm{\sigma}^k\right]\bm{\sigma}^l\right]\nn\\
&\quad\times\int d\Omega \hat{\bm{r}}_0^i\hat{\bm{r}}_0^{j}\int dr\,r^3 \psi_{0}^{{(m)}\dag}(r)\phi^{(n)}(r)=i\frac{2}{3}\left(\hat{e}_{m^{\prime}_J}^{*}\times \hat{e}_{m_J}\right)^j\int dr\,r^3 \psi_{0}^{{(m)}\dag}(r)\phi^{(n)}(r)\nn\\
&=i\frac{2}{3}\left(\hat{e}_{m^{\prime}_J}^{*}\times \hat{e}_{m_J}\right)^j\langle\Psi(m^3\mathcal{S}_1)|\Psi_{\la}^{\dag} \hat{\bm{r}}^{\dag}_{\la}\cdot\bm{r}S|S(n^3S_1)\rangle\,,
\end{align}
and due to the symmetry of the angular integral with respect to the indices [see \eq{agint}], we can also derive
\begin{align}
\langle\Psi(m^3\mathcal{S}_1)|\left(\bm{\sigma}_1+\bm{\sigma}_2\right)\cdot\bm{r}\hat{\bm{r}}_{\la}^{j\dag}|S(n^3S_1)\rangle=i\frac{2}{3}\left(\hat{e}_{m^{\prime}_J}^{*}\times \hat{e}_{m_J}\right)^j\langle\Psi(m^3\mathcal{S}_1)|\hat{\bm{r}}^{\dag}_{\la}\cdot\bm{r}|S(n^3S_1)\rangle\,.
\end{align}

\section{Hadronization of local gluonic operators }
\subsection{Two-pion production through the anomaly of the energy-momentum tensor}
\label{Sec:anomaly}

The traceless part of the energy momentum tensor for QCD is
\begin{align}
\theta_{\mu\nu}-\frac{1}{4}g_{\mu\nu}\theta^{\alpha}_{\alpha}=\frac{1}{4}g_{\mu\nu}G^{\alpha\beta a}G^a_{\alpha\beta}-G^a_{\mu\alpha}G^{\alpha\,a}_{\nu}+i\sum_i\bar{q}_i\gamma_{\mu}D_{\nu}q_i-\sum_i\frac{1}{4}g_{\mu\nu}m_{i}\bar{q}_iq_i
\,.
\end{align}
The trace of the energy momentum tensor reads
\begin{align}
\theta^{\mu}_{\mu}=\frac{1}{4}\frac{\beta(\alpha_s)}{\alpha_s}G^{\mu\nu a}G_{\mu\nu a}+\sum_i(1-\gamma_i)m_i\bar{q}_iq_i\,,\label{tremt}
\end{align}
where the $\gamma_i$ are the anomalous dimension of the $\bar{q}_iq_i$ operator and $\beta$ is the QCD $\beta$-function
\begin{align}
\beta(\alpha_s)=-\frac{\beta_0\alpha^2_s}{2\pi}+\mathcal{O}(\alpha^3_s)\,,\quad \beta_0=\frac{11N_c}{3}-\frac{4T_F n_f}{3}\,.
\end{align}
The first term in \eq{tremt} is generated by the conformal anomaly, and the second by the explicit breaking of the scale invariance due to 
the quark masses.

The matrix element $\langle \pi^+(p_+)\pi^-(p_-)|\theta_{\mu\nu}|0\rangle$ can be determined up to quadratic order in $p_+$, $p_-$ and $m_{\pi}$ from \cite{Voloshin:2007dx,Novikov:1980fa}
\begin{itemize}
\item Conservation on the mass shell: $(p_++p_-)^{\mu}\langle \pi^+(p_+)\pi^-(p_-)|\theta_{\mu\nu}|0\rangle_{p^2_+=p^2_-=m^2_{\pi}}=0$.
\item Normalization: $\langle \pi^+(p)\pi^-(-p)|\theta_{\mu\nu}|0\rangle_{p^2=m^2_{\pi}}=2p_{\mu}p_{\nu}$.
\item Adler zero condition: $\langle \pi^+(p)\pi^-(0)|\theta_{\mu\nu}|0\rangle_{p^2=m^2_{\pi}}=0$.
\end{itemize}
and reads as
\begin{align}
&\langle \pi^+(p_+)\pi^-(p_-)|\theta_{\mu\nu}|0\rangle=\left[(p_+\cdot p_-+p^2_++p^2_-)-m^2_{\pi}\right]g_{\mu\nu}-p_{+\mu}p_{-\nu}-p_{+\nu}p_{-\mu}\\
&\langle \pi^+(p_+)\pi^-(p_-)|\theta^{\mu}_{\mu}|0\rangle=2(p_+\cdot p_-)+4m^2_{\pi}\,.\label{idk1}
\end{align}

The square of chromoelectric and chromomagnetic fields can be written as
\begin{align}
\bm{E}^2&=v^{\mu}v^{\nu}\theta^g_{\mu\nu}-\frac{1}{4}G^{\alpha\beta a}G^a_{\alpha\beta}\,,\\
\bm{B}^2&=v^{\mu}v^{\nu}\theta^g_{\mu\nu}+\frac{1}{4}G^{\alpha\beta a}G^a_{\alpha\beta}\,,
\end{align}
with $v_{\mu}=(1,\,\bm{0})$ and 
\begin{align}
\theta^g_{\mu\nu}=\frac{1}{4}g_{\mu\nu}G^{\alpha\beta a}G^a_{\alpha\beta}-G^a_{\mu\alpha}G^{\alpha\,a}_{\nu}\,.
\end{align}
We can rewrite the chromoelectric and chromomagnetic fields in terms of $\theta^{\mu}_{\mu}$,
\begin{align}
\bm{E}^2&=v^{\mu}v^{\nu}\theta^g_{\mu\nu}-\frac{\alpha_s}{\beta(\alpha_s)}\left(\theta^{\mu}_{\mu}-\sum_i m_i(1-\gamma_i)\bar{q}_iq_i\right)\,,\\
\bm{B}^2&=v^{\mu}v^{\nu}\theta^g_{\mu\nu}+\frac{\alpha_s}{\beta(\alpha_s)}\left(\theta^{\mu}_{\mu}-\sum_i m_i(1-\gamma_i)\bar{q}_iq_i\right)\,.
\end{align}
The problem of computing the matrix elements $\langle \pi^+(p_+)\pi^-(p_-)|\bm{E}^2|0\rangle$ and $\langle \pi^+(p_+)\pi^-(p_-)|\bm{B}^2|0\rangle$ is transformed into the computation of $\langle \pi^+(p_+)\pi^-(p_-)|\theta^g_{\mu\nu}|0\rangle$, $\langle \pi^+(p_+)\pi^-(p_-)|\theta^{\mu}_{\mu}|0\rangle$, and $\langle \pi^+(p_+)\pi^-(p_-)|\sum_i m_i\bar{q}_iq_i|0\rangle$. These can be evaluated in $\chi$PT except for a possible normalization, and they correspond to the traceless part of the energy-momentum tensor, the trace of the energy-momentum tensor, and the scalar current, respectively. Let us start by the trace of the energy-momentum tensor
\begin{align}
\langle \pi^+(p_+)\pi^-(p_-)|\theta^{\mu}_{\mu}|0\rangle=&V_1(\mu)\left[-\frac{F^2}{2}\langle\pa_{\mu}U^{\dag}\pa^{\mu}U\rangle-F^2\langle\chi U^{\dag}+U\chi^{\dag}\rangle+\dots\right]
\nn
\\=&
V_1(\mu)\left[2(p_+\cdot p_-)+4m^2_{\pi}+\dots\right]\,.
\end{align}
If we compare with \eq{idk1}, we determine that $V_1(\mu)=1$. The matrix element of the scalar current reads
\begin{align}
\langle \pi^+(p_+)\pi^-(p_-)|\sum_i m_i\bar{q}_iq_i|0\rangle=-\frac{F^2}{4}\langle\chi U^{\dag}+U\chi^{\dag}\rangle+\dots=m^2_{\pi}+\dots\,,
\end{align}
where one can think the parameter $B$ ($\chi=2B\hat{m}\mathbbm{1}$) as taking the role of the normalization parameter. Finally, for $\theta^{g}_{\mu\nu}$, we have no way to determine the normalization
\begin{align}
\langle \pi^+(p_+)\pi^-(p_-)|\theta^{g}_{\mu\nu}|0\rangle&=V_2(\mu)\left[\frac{F^2}{4}\langle\pa_{\mu}U^{\dag}\pa_{\nu}U+\pa_{\mu}U^{\dag}\pa_{\nu}U-\frac{g_{\mu\nu}}{2}\pa^{\alpha}U^{\dag}\pa_{\alpha}U\rangle+\dots\right] \nn\\
&=-V_2(\mu)\left[p_{+\mu}p_{-\nu}+p_{-\mu}p_{+\nu}-\frac{1}{2}g_{\mu\nu}p_+\cdot p_-+\dots\right]\,,
\end{align}
Adding everything together we arrive at
\begin{align}
\langle \pi^+(p_+)\pi^-(p_-)|\frac{\beta_0\alpha_s}{2\pi}\bm{E}^2|0\rangle&=\left(2-\frac{9}{2}\kappa\right)p^0_+p^0_--\left(2+\frac{3}{2}\kappa\right)\bm{p}_+\cdot\bm{p}_-+3m^2_{\pi}\,,\label{eeh2p}\\
-\langle \pi^+(p_+)\pi^-(p_-)|\frac{\beta_0\alpha_s}{2\pi}\bm{B}^2|0\rangle&=\left(2+\frac{9}{2}\kappa\right)p^0_+p^0_--\left(2-\frac{3}{2}\kappa\right)\bm{p}_+\cdot\bm{p}_-+3m^2_{\pi}\,,\label{bbh2p}
\end{align}
where we have defined $\kappa=\alpha_s \beta_0 V_2(\mu)/(6\pi)$ as in Ref.~\cite{Novikov:1980fa}, and we arrive at the results of Refs.~\cite{Voloshin:2007dx,Brambilla:2015rqa}. We emphasize that in this relation we have neglected higher order $\alpha_s$ corrections to the beta function. These $\alpha_s$ corrections are computed at a low-energy scale, which makes their computation unfeasible. Still, they could be parametrized as an overall coefficient (redefining $\kappa$) of the equation. In \eq{bbh2p}, we have also neglected the anomalous dimension of the $\bar q q$ operator. Otherwise, the coefficient proportional to $m^2_{\pi}$ becomes arbitrary. Whereas this is customary done, this should be further investigated. 

\subsection{Single pion creation from the axial anomaly}\label{Sec:AxialAnomaly}

The axial current matrix elements are defined as
\begin{align}
\langle 0|\bar{q}(0)\gamma^{\mu}\gamma_5\frac{\lambda_i}{2}q(0)|\pi_j(p)\rangle=i Fp^{\mu}\de_{ij}\,,
\end{align}
with $\lambda$ the Gell-Mann matrices and $q=(u,\,d,\,s)$. The divergence of the axial current reads
\begin{align}
\pa_{\mu}(\bar{u}\gamma_{\mu}\gamma_5u)=2i m_u \bar{u}\gamma_5u+\frac{\alpha_s}{4\pi}G^a_{\mu\nu}\tilde{G}^{a,\mu\nu}\,,
\end{align}
with
\begin{align}
\tilde{G}^{\mu\nu}=\frac{1}{2}\epsilon^{\mu\nu\alpha\beta}G_{\alpha\beta} \qquad (\epsilon^{0123}=1)\,,
\end{align}
and the same holds replacing $u$ by $d$ or $s$. Then, we have
\begin{align}
\partial_{\mu}\langle 0|\bar{q}\gamma^{\mu}\gamma_5\frac{\lambda_3}{2}q|\pi^0\rangle&=\frac{1}{2}\partial_{\mu}\langle 0|\bar{u}\gamma^{\mu}\gamma_5u-\bar{d}\gamma^{\mu}\gamma_5d|\pi^0\rangle=i\langle 0|m_u\bar{u}\gamma_5u-m_d\bar{d}\gamma_5d|\pi^0\rangle\nn\\
&=\frac{i}{2}\langle 0|(m_u+m_d)(\bar{u}\gamma_5u-\bar{d}\gamma_5d)|\pi^0\rangle=Fm^2_{\pi^0}\,.
\label{acds1}
\end{align}
Since the diagonal flavor current does not interpolate the $\pi^0$ we have
\begin{align}
&0=\partial_{\mu}\langle 0|\bar{u}\gamma^{\mu}\gamma_5u+\bar{d}\gamma^{\mu}\gamma_5d|\pi^0\rangle=\langle 0|2im_u\bar{u}\gamma_5u+2im_d\bar{d}\gamma_5d+2\frac{\alpha_s}{4\pi}G_{\mu\nu}\tilde{G}^{\mu\nu}|\pi^0\rangle \nn\\
&=\langle 0|i(m_u-m_d)(\bar{u}\gamma_5u-\bar{d}\gamma_5d)+2\frac{\alpha_s}{4\pi}G_{\mu\nu}\tilde{G}^{\mu\nu}|\pi^0\rangle\,,\label{acds2}
\end{align}
and using \eq{acds1} in \eq{acds2}, we obtain
\begin{align}
\langle 0|\frac{\alpha_s}{2\pi}G_{\mu\nu}\tilde{G}^{\mu\nu}|\pi^0\rangle=-i(m_u-m_d)\langle 0|\bar{u}\gamma_5u-\bar{d}\gamma_5d|\pi^0\rangle=2\frac{m_d-m_u}{m_u+m_d}Fm^2_{\pi^0}\,,\label{acds3}
\end{align}
which is the result from Ref.~\cite{Gross:1979ur}. Finally, we note that
\begin{align}
\langle 0|\frac{\alpha_s}{\pi}\bm{E}\cdot\bm{B}|\pi^0\rangle=\frac{1}{4}\langle 0|\frac{\alpha_s}{\pi}G_{\mu\nu}\tilde{G}^{\mu\nu}|\pi^0\rangle=\frac{m_d-m_u}{m_u+m_d}F m^2_{\pi^0}\,.
\end{align}
A similar analysis for the $\eta$ produces
\begin{align}
\langle 0|\frac{3\alpha_s}{4\pi}G_{\mu\nu}\tilde{G}^{\mu\nu}|\eta \rangle=\sqrt{3}F m^2_{\eta}\,.
\end{align}
We remark that, compared with the prediction of the previous section, these results do not have neither ${\cal O}(\alpha_s)$ nor ${\cal O}(p^4)$ corrections. 

\subsection{Single \texorpdfstring{$P$}{P}-wave pion anomaly}\label{Sec:PwaveAnomaly}

We first consider the matrix elements of following gluonic operators
\begin{align}
&i\langle\pi^0(p)|G^a_{\mu\nu}\left(D_{\rho}G_{\lambda\sigma}\right)^a|0\rangle=X p_{\rho}\epsilon_{\mu\nu\lambda\sigma}+Y\left(p_{\lambda}\epsilon_{\mu\nu\rho\sigma}-p_{\sigma}\epsilon_{\mu\nu\rho\lambda}\right)\,, \label{idk3}\\
&i\langle\pi^0(p)|\left(D_{\rho}G_{\mu\nu}\right)^aG^a_{\lambda\sigma}|0\rangle=X^{\prime}p_{\rho}\epsilon_{\mu\nu\lambda\sigma}+Y^{\prime}\left(p_{\mu}\epsilon_{\lambda\sigma\rho\nu}-p_{\nu}\epsilon_{\lambda\sigma\rho\mu}\right)\,,\label{idk2}
\end{align}
where we have used the Schouten identity
\begin{align}
p_{\rho}\epsilon_{\mu\nu\lambda\sigma}=p_{\lambda}\epsilon_{\mu\nu\rho\sigma}-p_{\sigma}\epsilon_{\mu\nu\rho\lambda}-p_{\mu}\epsilon_{\nu\rho\lambda\sigma}+p_{\nu}\epsilon_{\mu\rho\lambda\sigma}\,.\label{mridk}
\end{align}
Due to the Jacobi identity
\begin{align}
D_{\rho}G_{\lambda\sigma}+D_{\sigma}G_{\rho\lambda}+D_{\lambda}G_{\sigma\rho}=0\,,
\end{align}
 we have
\begin{align}
0&=i\langle\pi^0(p)|G^a_{\mu\nu}\left(D_{\rho}G_{\lambda\sigma}\right)^a|0\rangle+i\langle\pi^0(p)|G^a_{\mu\nu}\left(D_{\sigma}G_{\rho\lambda}\right)^a|0\rangle+i\langle\pi^0(p)|G^a_{\mu\nu}\left(D_{\lambda}G_{\sigma\rho}\right)^a|0\rangle\nonumber\\
&=X\left(p_{\rho}\epsilon_{\mu\nu\lambda\sigma}+p_{\sigma}\epsilon_{\mu\nu\rho\lambda}+p_{\lambda}\epsilon_{\mu\nu\sigma\rho}\right)+Y\left(p_{\lambda}\epsilon_{\mu\nu\rho\sigma}+p_{\rho}\epsilon_{\mu\nu\sigma\lambda}+p_{\sigma}\epsilon_{\mu\nu\lambda\rho}\right)
\nn
\\
&\quad
-Y\left(p_{\sigma}\epsilon_{\mu\nu\rho\lambda}+p_{\lambda}\epsilon_{\mu\nu\sigma\rho}+p_{\rho}\epsilon_{\mu\nu\lambda\sigma}\right)\nonumber\\
&=\left(X-2Y\right)\left(p_{\rho}\epsilon_{\mu\nu\lambda\sigma}+p_{\sigma}\epsilon_{\mu\nu\rho\lambda}+p_{\lambda}\epsilon_{\mu\nu\sigma\rho}\right)\Rightarrow\,X=2Y\,,
\end{align}
and similarly $X^{\prime}=2Y^{\prime}$.

The sum of the two matrix elements is proportional to the pion momentum:
\begin{align}
&i\langle\pi^0(p)|G^a_{\mu\nu}\left(D_{\rho}G_{\lambda\sigma}\right)^a|0\rangle+i\langle\pi^0(p)|\left(D_{\rho}G_{\mu\nu}\right)^aG^a_{\lambda\sigma}|0\rangle \nonumber\\
=&i\langle\pi^0(p)|\partial_{\rho}\left(G^a_{\mu\nu}G^a_{\lambda\sigma}\right)|0\rangle +\langle\pi^0(p)|f^{aeb}gA^b_{\rho}G^e_{\mu\nu}G^a_{\lambda\sigma}+G^a_{\mu\nu}f^{acb}gA^b_{\rho}G^c_{\lambda\sigma}|0\rangle\nonumber\\
=&i\int d^4xe^{-ip \cdot x}\langle 0|\pi^0(x)\partial_{\rho}\left(G^a_{\mu\nu}G^a_{\lambda\sigma}\right)|0\rangle=-p_{\rho}\langle\pi^0(p)|G^a_{\mu\nu}G_{\lambda\sigma}^a|0\rangle\,.\label{idk5}
\end{align}
The sum of the matrix elements in Eqs.~\eqref{idk3}-\eqref{idk2} is
\begin{align}
&i\langle\pi^0(p)|G^a_{\mu\nu}\left(D_{\rho}G_{\lambda\sigma}\right)^a|0\rangle+i\langle\pi^0(p)|\left(D_{\rho}G_{\mu\nu}\right)^aG^a_{\lambda\sigma}|0\rangle \nn\\
&=\left(X+X^{\prime}+Y^{\prime}\right) p_{\rho}\epsilon_{\mu\nu\lambda\sigma}+(Y-Y^{\prime})\left(p_{\lambda}\epsilon_{\mu\nu\rho\sigma}-p_{\sigma}\epsilon_{\mu\nu\rho\lambda}\right)\,,\label{idk4}
\end{align}
where we have used \eq{mridk} to reexpress \eq{idk2}. In order for \eq{idk4} to be compatible with \eq{idk5}, the relation $Y=Y^{\prime}$ must hold. Therefore, we arrive at
\begin{align}
&i\langle\pi^0(p)|G^a_{\mu\nu}\left(D_{\rho}G_{\lambda\sigma}\right)^a|0\rangle+i\langle\pi^0(p)|\left(D_{\rho}G_{\mu\nu}\right)^aG^a_{\lambda\sigma}|0\rangle=\frac{5}{2}Xp_{\rho}\epsilon_{\mu\nu\lambda\sigma}\nn\\
&=-p_{\rho}\langle\pi^0(p)|G^a_{\mu\nu}G_{\lambda\sigma}^a|0\rangle\,.
\end{align}
Contracting both sides of the last equality with $\epsilon^{\mu\nu\lambda\sigma}/2$,
\begin{align}
 X=-\frac{1}{30}\langle\pi^0(p)|G^a_{\mu\nu}\tilde{G}^{\mu\nu a}|0\rangle\,,
\end{align}
the matrix element is precisely the one given in \eq{acds3}.

Now we can write the matrix elements we are interested in,
\begin{align}
&i\langle\pi^0(p)|\bm{E}^a_{i}\left(\bm{D}_{j}\bm{B}_{k}\right)^a|0\rangle=\frac{i}{2}\epsilon^{0klm}\langle\pi^0(p)|G^a_{i0}\left(D_{j}G_{lm}\right)^a|0\rangle=-\frac{1}{2}X\left[3p_j\delta_{ki}-p_i\delta_{kj}\right]\,,\\
&i\langle\pi^0(p)|\left(\bm{D}_{j}\bm{B}_{k}\right)^a\bm{E}^a_{i}|0\rangle=\frac{i}{2}\epsilon^{0klm}\langle\pi^0(p)|\left(D_{j}G_{lm}\right)^aG^a_{i0}|0\rangle=-\frac{1}{2}X\left[3p_j\delta_{ki}-p_i\delta_{kj}\right]\,.
\end{align}
We note that
\begin{align}
i\langle\pi^0(p)|\bm{E}^a_{i}\left(D_{j}\bm{B}_{k}\right)^a+\left(\bm{D}_{j}\bm{B}_{k}\right)^a\bm{E}^a_{i}|0\rangle=-X\left[3p_j\delta_{ki}-p_i\delta_{kj}\right]\,,
\end{align}
differs from Refs.~\cite{Voloshin:2003kn,Voloshin:2007dx} by a factor $1/2$.

\bibliographystyle{apsrev4-1}
\bibliography{biblio}

\end{document}